\documentclass[lettersize,journal]{IEEEtran}
\usepackage{times}
\usepackage{graphicx}
\usepackage{amsmath, amsfonts}
\usepackage{amssymb}
\usepackage{algorithm}
\usepackage{algorithmic}
\usepackage{multirow}
\usepackage{bm}
\usepackage{float}
\usepackage{lettrine}
\usepackage{stfloats}
\usepackage{cite}
\usepackage{amsmath}
\usepackage{diagbox}
\usepackage{makecell}
\usepackage{subfigure}
\usepackage{caption}
\usepackage{subfig} 
\usepackage{xcolor}
\usepackage{url}
\usepackage{nomencl}
\makenomenclature

\usepackage{enumitem}
\usepackage{booktabs}
\usepackage{tabularx}
\usepackage{xspace}

\newcommand{\eg}{\emph{e.g.,}\xspace}
\newcommand{\ie}{\emph{i.e.,}\xspace}

\begin{document}

\title{Democratic Recommendation with User and Item Representatives Produced by Graph Condensation}

\author{Jiahao Liang\IEEEauthorrefmark{2}, Haoran~Yang\IEEEauthorrefmark{2}, Xiangyu Zhao\IEEEauthorrefmark{1}, \IEEEmembership{Member,~IEEE,} 
Zhiwen~Yu, \IEEEmembership{Senior Member,~IEEE},\\
Guandong Xu, \IEEEmembership{Member,~IEEE,}
Wanyu Wang,
% C. L. Philip Chen, \IEEEmembership{Fellow,~IEEE}, 
Kaixiang Yang\IEEEauthorrefmark{1}, \IEEEmembership{Member,~IEEE}
\IEEEcompsocitemizethanks{
% \IEEEcompsocthanksitem Kaixiang Yang is with the State Key
% Laboratory of Industrial Control Technology,  College of Control Science and Engineering in Zhejiang University, China. Email: yangkaixiang@zju.edu.cn.
% \IEEEcompsocthanksitem Zhiwen Yu, Yuchen Liu and C. L. Philip Chen are with the School of Computer Science and Engineering in South China University of Technology, China.  Email: zhwyu@scut.edu.cn.(Corresponding author: Zhiwen Yu.)
\IEEEcompsocthanksitem Jiahao Liang, Kaixiang Yang are with the School of Computer Science and Engineering in South China University of Technology, China.(Corresponding author: Kaixiang Yang, e-mail: csjiahliang6@mail.scut.edu.cn, yangkx@scut.edu.cn)
\IEEEcompsocthanksitem Haoran Yang is with School of Computer Science and Engineering at Central South University, China. (e-mail: yanghaoran.1998@outlook.com.)
\IEEEcompsocthanksitem Xiangyu Zhao is with City University of Hong Kong, Department of Data Science, Hong Kong. (Corresponding author: Xiangyu Zhao, e-mail: xy.zhao@cityu.edu.hk.)
\IEEEcompsocthanksitem Zhiwen Yu are with the School of Computer Science and Engineering in South China University of Technology and the Pengcheng Lab, China. ( e-mail:zhwyu@scut.edu.cn.)
\IEEEcompsocthanksitem Guandong Xu is with Education University of Hong Kong, University Research Facility of Data Science and Artificial Intelligence. (e-mail: gdxu@eduhk.hk.)
% \IEEEcompsocthanksitem Xiaoqing Liu is with the School of Future Technology in South China University of Technology, Guangzhou 510641, and also with Pengcheng Laboratory, Shenzhen 518000, China (e-mail: ft\_liuxiaoqing@mail.scut.edu.cn).
\IEEEcompsocthanksitem Wanyu Wang is with City University of Hong Kong, Department of Information Systems, Hong Kong. (e-mail: wanyuwang4-c@my.cityu.edu.hk.)
% \IEEEcompsocthanksitem C.L.Philip Chen is with the School of Computer Science and Engineering in South China University of Technology and the Pazhou Lab, China (e-mail: philip.chen@ieee.org.)
}

% <-this % stops a space
\thanks{\IEEEauthorrefmark{1} indicates the corresponding authors. \IEEEauthorrefmark{2} indicates the equal contribution.}}

\markboth{IEEE Transactions on Knowledge and Data Engineering}
{Shell \MakeLowercase{\textit{et al.}}: IEEE Transactions on Knowledge and Data Engineering}

\IEEEcompsoctitleabstractindextext{

\begin{abstract}
%\yhr{(i) Improve the overall typesetting, especially pay attention to the end of each paragraph, no less than 4 words. (ii) Improve the writing quality by inputting every paragraph to ChatGPT for refining. These two things should be done when you complete all the other revisions.}
The challenges associated with large-scale user-item interaction graphs have attracted increasing attention in graph-based recommendation systems, primarily due to computational inefficiencies and inadequate information propagation. Existing methods provide partial solutions but suffer from notable limitations: model-centric approaches, such as sampling and aggregation, often struggle with generalization, while data-centric techniques, including graph sparsification and coarsening, lead to information loss and ineffective handling of bipartite graph structures. Recent advances in graph condensation offer a promising direction by reducing graph size while preserving essential information, presenting a novel approach to mitigating these challenges. Inspired by the principles of democracy, we propose \textbf{DemoRec}, a framework that leverages graph condensation to generate user and item representatives for recommendation tasks. By constructing a compact interaction graph and clustering nodes with shared characteristics from the original graph, DemoRec significantly reduces graph size and computational complexity. Furthermore, it mitigates the over-reliance on high-order information, a critical challenge in large-scale bipartite graphs. Extensive experiments conducted on four public datasets demonstrate the effectiveness of DemoRec, showcasing substantial improvements in recommendation performance, computational efficiency, and robustness compared to SOTA methods.
\end{abstract}

}
\maketitle
\IEEEdisplaynotcompsoctitleabstractindextext

\IEEEpeerreviewmaketitle

\section{Introduction}
\IEEEPARstart{G}{raph}-based recommender systems~\cite{wu2021self, chen2023heterogeneous, yu2022graph, cai2023lightgcl, peng2023dual} have become essential in various domains, including e-commerce~\cite{zhao2022joint,xv2023commerce}, social networks~\cite{fan2019graph,guo2020deep}, and content streaming~\cite{rappaz2021recommendation}, where personalized recommendations heavily depend on modeling user-item interactions. These interactions are typically represented as graphs, enabling systems to capture the structural relationships between users and items. However, the exponential growth of user-item interaction data has led to large-scale graphs that introduce significant computational and structural challenges~\cite{mao2021ultragcn,peng2022less}. This scalability bottleneck not only slows down recommendation pipelines but also complicates the maintenance of real-time performance, a critical requirement in many real-world applications.

Graph Neural Networks (GNNs) have emerged as a powerful tool to address some of these issues, demonstrating significant promise in learning rich and effective representations from graph-structured data~\cite{wang2020disentangled, yang2021graphformers}. By iteratively aggregating information from neighboring nodes, GNNs can generate embeddings that encapsulate both local and global structural features, thereby improving the quality of recommendations. Despite their strengths, GNNs are not immune to the challenges posed by large graphs. Their scalability and efficiency often falter as graph size increases, primarily due to the computational overhead of message passing across vast networks~\cite{chiang2019cluster, zheng2024structure}. This process, while effective for smaller graphs, becomes prohibitively expensive in terms of time and resources when applied to datasets with millions or billions of interactions, limiting the practical deployment of GNN-based systems in large-scale settings.

Moreover, user-item interaction graphs, which are typically structured as bipartite graphs, introduce additional complexities that further complicate recommendation tasks. In a bipartite graph, nodes are divided into two distinct sets, users and items, with edges existing only between the sets and not within them. This structure inherently restricts information propagation, as there are no direct user to user or item to item connections to facilitate the flow of information~\cite{li2021bipartite, he2016birank}. As a result, the embeddings learned by GNNs may fail to fully capture the nuanced relationships and dependencies across the graph, leading to suboptimal recommendation performance. This limitation is particularly pronounced in sparse graphs, where the scarcity of interactions exacerbates the challenge of propagating meaningful signals across the bipartite structure. These issues of scalability and efficiency in graph-based recommender systems, particularly in handling large-scale user-item interaction graphs, can be roughly categorized into two key aspects:
\begin{itemize}[leftmargin=*]
    \item[(i)] \textbf{The efficiency of processing large-scale user-item interaction graphs remains a critical issue.} The computational complexity of handling large-scale user-item interaction graphs remains a significant challenge. Traditional graph neural networks (GNNs) rely on message-passing mechanisms to propagate information across nodes, leading to substantial computational overhead. Early model-centric approaches~\cite{liu2021sampling,huang2018adaptive,he2024polarized, bi2024graph, bi2023two}, such as sampling-based methods and linear aggregation~\cite{gong2023gsampler,tian2008efficient,dong2022improving}, attempt to alleviate this burden by sampling subsets of neighbors or precomputing message aggregations. However, their generalization ability is often limited, as their effectiveness depends heavily on architectural design and hyperparameter selection~\cite{zhang2024training}. In contrast, data-centric approaches like graph sparsification and coarsening~\cite{bravo2019unifying,cai2021graph,li2022graph} aim to reduce graph size by eliminating redundant edges or merging similar nodes. While these methods improve storage efficiency and reduce computational costs, they frequently lead to information loss~\cite{hashemi2024comprehensive}, ultimately diminishing recommendation accuracy. Moreover, they heavily rely on heuristic techniques, such as principal eigenvalues or pairwise distances, which often fail to generalize across different datasets and recommendation tasks. A critical limitation of existing methods is their lack of optimization for bipartite graphs, resulting in suboptimal information propagation.
    \item [(ii)] \textbf{The issue of insufficient propagation of information in bipartite graphs further complicates the design of recommender systems.} The structural characteristics of bipartite graphs introduce additional challenges for recommender systems. Unlike general graphs, bipartite graphs do not contain direct edges within the same node set (\eg user-user or item-item connections), restricting effective information propagation. To mitigate this issue, many approaches incorporate high-order information by stacking multiple GNN layers, allowing information to flow across longer paths in the graph. However, this strategy significantly increases computational costs and exacerbates scalability issues when handling large graphs. Furthermore, deeper architectures often suffer from over-smoothing, where node representations become indistinguishable, ultimately degrading model performance \cite{hashemi2024comprehensive}.
\end{itemize}

\begin{figure}[h]
	\centering
	\includegraphics[width=1\linewidth]{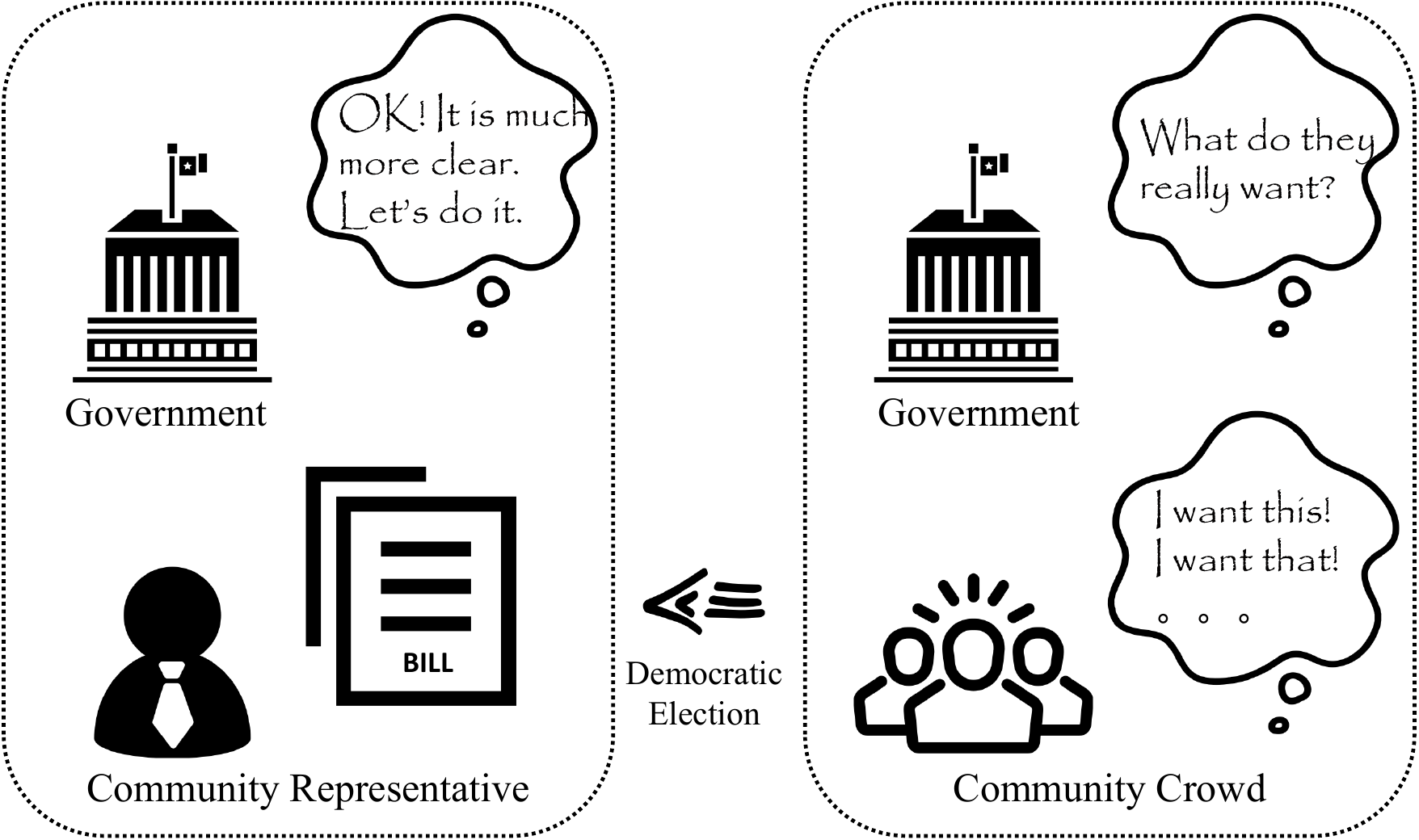}
	\caption{A toy example of the democratic election process.}
	\label{fig:democratic_example}
\end{figure}

To better illustrate the motivation and rationale behind our solution, we draw an analogy to the democratic election process, as depicted in Fig.~\ref{fig:democratic_example}. In a community, individuals elect representatives to voice their collective opinions in decision-making. This democratic process enhances efficiency by reducing the number of participants while preserving the key demands and characteristics of the larger population. By filtering out unnecessary noise, it facilitates more effective information propagation and decision-making.
Recent advances~\cite{jin2021graph,gao2024graph} in graph condensation offer a promising method to achieve such a democratic process in graph-based recommendation scenarios to address aforementioned challenges. Graph condensation techniques aim to create compact representations of graphs by generating representative nodes that hopefully aggregate the shared characteristics of original nodes. This approach not only reduces graph size but also preserves the essential structural and feature information for downstream tasks. %\yhr{The challenge of implementing graph condensation should not appear before the story.}

However, directly applying graph condensation to recommendation scenarios is non-trivial due to the unique challenges posed by bipartite graph structures. To address this issue, we propose DemoRec, a novel framework that employs graph condensation to generate user and item representatives. In the context of recommendation, DemoRec identifies representative users and items within large-scale graphs, effectively reducing high-order information propagation costs while filtering out noise and preserving core features essential for accurate recommendations. Unlike conventional graph condensation techniques that focus on general graph structures, our approach is specifically designed for bipartite graphs, which are prevalent in recommendation systems. We introduce bipartite structure constraints to ensure that the condensed graph maintains essential user-item relationships. Furthermore, we propose the Bipartite Structure Loss (BSL), which enforces structural consistency by penalizing deviations from the original bipartite graph's connectivity patterns. BSL ensures that the condensed graph retains critical user-item interaction properties, with theoretical guarantees of its effectiveness and ablation studies demonstrating its significant contribution to recommendation accuracy. This refinement not only overcomes the limitations of directly applying graph condensation but also represents a significant methodological innovation by adapting the technique to meet the unique requirements in scenarios of graph-based recommender systems.

%Inspired by these advancements, we propose a novel framework, Democratic Recommendation with User and Item Representatives Produced by Graph Condensation (DemoRec).

%\yhr{This part is too short, three aspects of content should be added: (i) Tell a story about the democratic process with toy examples, explaining the rationale of such a strategy. (ii) Emphasizing that directly adopting graph condensation is non-trivial by mentioning the proposed bipartite structure constraints. (iii) Simple introduction to the overall framework.} DemoRec addresses the efficiency and propagation challenges in large-scale bipartite graphs by generating user and item representatives that effectively capture the core properties of the graph while eliminating the need for high-order information propagation.

The contributions of this work can be summarized into three points, which are as follows:
\begin{itemize}[leftmargin=*]
\item We introduce DemoRec, the first graph condensation-based framework for recommender systems, which leverages user and item representatives to tackle the scalability and propagation challenges of large-scale bipartite graphs.
\item DemoRec eliminates reliance on high-order information, thereby reducing computational complexity while retaining essential structural properties of the original graph.
\item We introduce BSL, which enforces structural consistency in the condensed graph by preserving bipartite relationships, with theoretical proof guaranteeing its effectiveness and ablation studies empirically validating its contribution to recommendation performance.
\item We conduct five extensive experiments on benchmark datasets to demonstrate the superior performance of DemoRec in terms of scalability, recommendation accuracy, and generalization compared to existing state-of-the-art methods.
\end{itemize}

\section{Problem Statement}\label{sec:problem}
In this section, we formally define two core problems that underpin our research: Graph Condensation and Graph Recommendation. These problem statements establish the theoretical framework and objectives that guide our work, providing clarity on the challenges we aim to address and the goals we seek to achieve.

\subsection{Graph Condensation.}
Given a large, original graph $\mathcal{G}_{tr} = (\mathcal{V}_{tr}, \mathcal{A}_{tr}, \mathbf{X}_{tr})$, where $\mathcal{A}_{tr}$ and $\mathbf{X}_{tr}$ are the adjacency and feature matrices, the objective of graph condensation is to synthesize a much smaller condensed graph $\mathcal{G}_{c} = (\mathcal{V}_{c}, \mathcal{A}_{c}, \mathbf{X}_{c})$. This condensed graph must satisfy the size constraint $|\mathcal{V}_c| \ll |\mathcal{V}_{tr}|$ while retaining the essential information from $\mathcal{G}_{tr}$ for downstream tasks.

The parameters of this condensed graph, specifically its structure $\mathcal{A}_c$ and features $\mathbf{X}_c$, are the targets of our optimization. To achieve this, the condensation process is structured into three interdependent stages:

\begin{itemize}
    \item \textbf{Initialization:} The process begins by initializing the learnable parameters of the small graph, $\mathcal{A}_c$ and $\mathbf{X}_c$. This step is necessary to create the optimization starting point.
    
    \item \textbf{Relay Model Training:} To measure the "information" captured in $\mathcal{G}_c$, we introduce a \textbf{relay model}, denoted as a GNN $f_\theta$, and a \textbf{relay task}, $\mathcal{L}_{task}$ (\eg node classification or link prediction). The purpose of this model is to act as a surrogate, providing a learning signal. This model is trained on both the original graph $\mathcal{G}_{tr}$ and the condensed graph $\mathcal{G}_c$ to capture their respective learning dynamics.

    \item \textbf{Graph Alignment:} The core objective is to ensure that the relay model $f_\theta$, when trained on $\mathcal{G}_c$, behaves similarly to when it is trained on $\mathcal{G}_{tr}$. This similarity is enforced by minimizing an alignment loss, $\mathcal{L}_{match}$, which measures the discrepancy between the learning signals from the two graphs.
\end{itemize}

Formally, we define the relay task loss on the original and condensed graphs as:
\begin{equation}
    \mathcal{L}_{tr} = \mathcal{L}_{task}(f_\theta(\mathcal{A}_{tr}, \mathbf{X}_{tr}))
\end{equation}
\begin{equation}
    \mathcal{L}_{c} = \mathcal{L}_{task}(f_\theta(\mathcal{A}_{c}, \mathbf{X}_{c}))
\end{equation}
The goal of graph condensation is to find the optimal condensed graph parameters ($\mathcal{A}_c^*, \mathbf{X}_c^*$) by solving the following optimization problem, typically by matching the gradients produced by the relay task:
\begin{equation}
    (\mathcal{A}_c^*, \mathbf{X}_c^*) = \arg\min_{\mathcal{A}_c, \mathbf{X}_c} \mathbb{E}_{\theta \sim p(\theta)} [\mathcal{L}_{match}(\nabla_{\theta} \mathcal{L}_{tr}, \nabla_{\theta} \mathcal{L}_{c})]
\end{equation}
where $\mathcal{L}_{match}$ is a distance function (\eg cosine distance) between the gradients. By iteratively optimizing $\mathcal{A}_c$ and $\mathbf{X}_c$ to minimize this alignment loss, $\mathcal{G}_c$ is progressively refined to distill the critical information from $\mathcal{G}_{tr}$.

\subsection{Graph Recommendation.}
In the domain of recommender systems, the Graph Recommendation aims to predict Top-K item recommendations by modeling user-item interactions as a bipartite graph, $\mathcal{G} = (\mathcal{V}, \mathcal{A}, \mathbf{X})$. Here, $\mathcal{V} = \mathcal{V}_U \cup \mathcal{V}_I$ are user and item nodes, $\mathcal{A}$ is the sparse adjacency matrix, and $\mathbf{X}$ is the node feature matrix.Graph Neural Networks (GNNs) serve as powerful encoders, $f_\theta$, to learn node embeddings $\mathbf{Z} \in \mathbb{R}^{|\mathcal{V}| \times d}$ that capture preferences and relationships:
\begin{equation}\mathbf{Z} = f_\theta(\mathcal{A}, \mathbf{X}).
\end{equation}
The GNN is typically pre-trained on a source graph $\mathcal{G}_{tr}$ by optimizing its parameters $\theta$ via a standard recommendation loss $\mathcal{L}_{rec}$ (\eg BPR loss):
\begin{equation}\theta^* = \arg\min_{\theta} \mathcal{L}_{rec}(f_\theta(\mathcal{A}_{tr}, \mathbf{X}_{tr})).
\end{equation}
The trained model $f_{\theta^*}$ is then used to rank items and generate recommendations.

% In our framework, we employ Graph Condensation as a preliminary step to enhance the recommendation process. Let \( U = \{ u_1, \ldots, u_n, \ldots, u_N \} \) and \( I = \{ i_1, \ldots, i_m, \ldots, i_M \} \) represent the sets of users and items, respectively, where \( n \) and \( m \) denote the \( n \)-th user and the \( m \)-th item. We denote the interactions of the \( n \)-th user as \( S_n = \{ i_1, i_2, \ldots, i_t \} \), where each \( i \) corresponds to an item interacted with by the user.

% Our objective is to apply Graph Condensation to create a condensed representation of user-item interactions, which allows for more efficient processing in the recommendation task. Specifically, after obtaining the condensed graph, we aim to predict which item the user is likely to engage with next. Given the user-item interactions up to an arbitrary time \( t \), our goal is to accurately forecast the next item at time \( t+1 \) based on the insights derived from the condensed graph.

\section{Framework}
\label{sec:framework}
%\vspace{1mm}
% Implementing the proposed framework presents significant challenges. First, the current literature lacks research on Graph Condensation (GC) specifically applied to recommendation scenarios, providing limited guidance for implementing GC in Recommender Systems(RS). Second, existing methods treat nodes in the User-Item (U-I) graph uniformly. However, in the recommendation context, users and items require distinct handling approaches, necessitating adaptations to current methods before their application to recommendation tasks. Third, the optimization process involves substantial computational parameters due to the separate optimization of the bipartite user-item graph, and there are no inherent constraints ensuring that the new user and item representations maintain a bipartite structure.

In this section, we introduce our proposed GC framework for RS, termed DemoRec. In this framework, the user and item representatives refer to the condensed user and item nodes generated through the GC process. We first present the overall architecture of DemoRec, demonstrating how it effectively addresses the previously mentioned challenges. Finally, we summarize the overall model optimization process with an algorithm table.

\subsection{Overall Architecture}
\label{sec:architecture}

\begin{figure*}[t]
	\includegraphics[width=1\textwidth]{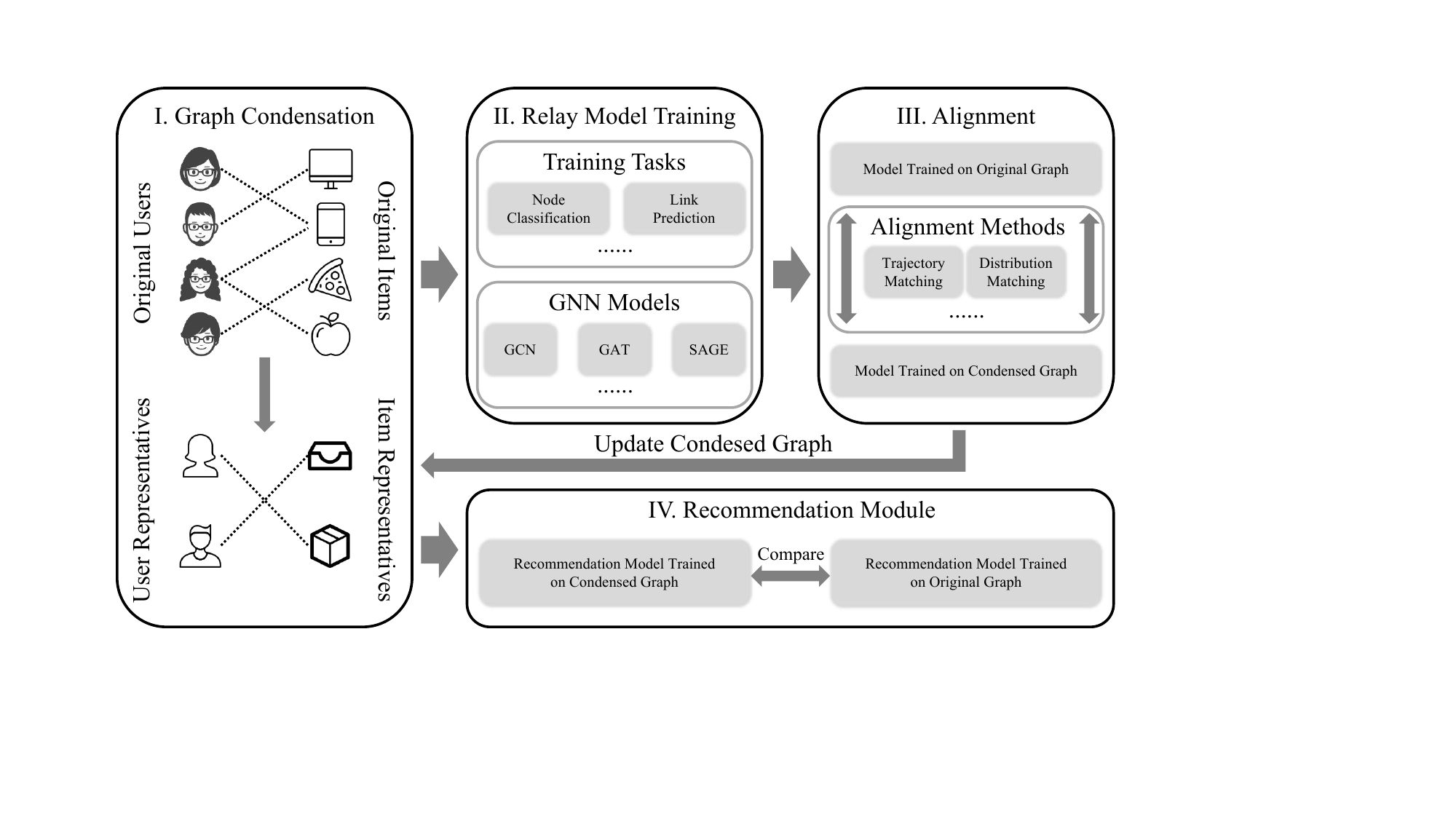}
        % \vspace{-20mm}
	\caption{The detailed architecture of the proposed DemoRec. The architecture consists of two distinct parts: the Graph Condensation module and the Recommendation module.}
	\label{fig:Fig1_Overall}
% 	\vspace{-5mm}
\end{figure*}

In this paper, we present the DemoRec framework, a novel two-stage graph recommendation framework that first conducts GC to produce the condensed user-item interaction graph and then performs recommendation tasks. Training on the condensed graph significantly improves training efficiency while maintaining essential properties of the original graph, enabling faster and more scalable recommendations. Additionally, the framework enhances information propagation by effectively preserving structural and semantic information from the original graph, which facilitates better recommendation quality.

As illustrated in Fig.~\ref{fig:Fig1_Overall}, our framework consists of two main stages: the Graph Condensation Stage (Stage One) and the Recommendation Stage (Stage Two), as illustrated in Fig.~\ref{fig:Fig1_Overall}. Stage One, the Graph Condensation Stage, aims to generate an optimized condensed graph that retains the essential structural and semantic information of the original user-item interaction graph. It includes three key modules: Module I (Condensed Graph Initialization), Module II (Relay Model Training), and Module III (Graph Alignment), which work iteratively to achieve graph condensation. Stage Two, the Recommendation Stage, represented by Module IV (Recommendation Module), utilizes the condensed graph for downstream recommendation tasks. These two stages are interconnected, with the output of the Graph Condensation Stage directly serving as the input for the Recommendation Stage. Together, they form a unified framework for efficient and scalable graph-based recommendations.

Stage One, the Graph Condensation Stage, aims to generate and optimize a condensed graph that preserves the essential structural and semantic information of the original user-item interaction graph through three interconnected modules. As shown in the leftmost part of Fig.~\ref{fig:Fig1_Overall}, Module I takes the original user-item interaction graph as input and generates a condensed graph with significantly fewer nodes, transforming the original user set $\boldsymbol{U}$ and item set $I$ into condensed representatives $\boldsymbol{U}'$ and $I'$, respectively. The Module I denotes the condensed graph initialization. Taking the original graph as the input and generating the condensed graph is the goal of all the three modules.
Module II, the Relay Model Training component, incorporates various Graph Neural Network (GNN) architectures (\eg GCN, GAT, and SAGE) to perform different training tasks such as node classification and link prediction, ensuring the condensed graph preserves essential structural properties. The framework is designed to be compatible with any of these backbones, allowing selection based on the specific task requirements. A detailed comparison of their performance as relay models is provided in our Component Analysis (Section \ref{sec:component}). Furthermore, the specific hyperparameters used for each GNN backbone, including their respective learning rates and \textbf{layer-wise configurations} (\eg, number of layers, hidden dimensions, and GAT heads), are detailed in our experimental setup (Section \ref{sec:Implementation Details}).
% We select one type of GNN as our component. 
The direct purpose of the relay model training is to provide a proxy to align the original graph and the condensed one in the subsequent module. Preserving essential structural properties is the goal of the entire GC part.
In Module III, we align the original and condensed graphs by implementing the alignment algorithm, such as trajectory matching and distribution matching, during the relay model training. The alignment ensures that the condensed graph retains the structural and semantic integrity of the original graph. These three modules work iteratively, with the condensed graph being continuously updated based on the alignment feedback to achieve optimal condensation.

Stage Two is the Recommendation Stage. The condensed graph optimized in Stage One is fed into Module IV, shown in the bottom part of Fig.~\ref{fig:Fig1_Overall}, to perform recommendation tasks. This stage leverages the condensed graph’s preserved structural and semantic properties to deliver high-quality recommendations efficiently.

% Upon obtaining the optimized condensed graph, the recommendation module, shown in the bottom part of Figure~\ref{fig:Fig1_Overall}, will process the condensed graph for the downstream recommendation task.

The subsequent contents will introduce the technical details of each module mentioned above.

\subsubsection{\textbf{Graph Condensation Stage}}
\label{sec:Graph condensation}
In order to solve the problem of insufficient information dissemination of bipartite graph, and avoid adding higher-order information, we introduce three key components of our GC framework. First, in the initial GC stage, we construct a bipartite graph $\mathcal{G} = (\mathcal{U}, \mathcal{I}, \mathcal{E})$ using the original user-item interaction dataset, where $\mathcal{U}$ and $\mathcal{I}$ represent the sets of users and items, respectively, and $\mathcal{E}$ denotes their interactions. Our goal is to generate a smaller condensed graph $\mathcal{G}' = (\mathcal{U}', \mathcal{I}', \mathcal{E}')$, where $|\mathcal{U}'| \ll |\mathcal{U}|$ and $|\mathcal{I}'| \ll |\mathcal{I}|$ while preserving key properties of the original graph. To achieve this, our framework operates on a central objective: we optimize the synthetic graph $\mathcal{G}'$ by ensuring that a GNN trained on $\mathcal{G}'$ (the 'relay model') produces a similar outcome (\eg training gradients or parameter trajectories) as the same GNN trained on the original graph $\mathcal{G}$. This "behavioral alignment" objective, formally defined later (\eg Eq. \ref{eq13}), guides the entire three-component process described below.

\paragraph{Initialization}\label{sec:Graph Initialization}
To initialize the condensed graph, we first create learnable embedding tables $\mathbf{E}_\mathcal{U} \in \mathbb{R}^{|\mathcal{U}'| \times d}$ and $\mathbf{E}_\mathcal{I} \in \mathbb{R}^{|\mathcal{I}'| \times d}$ for condensed users and items respectively, where $d$ is the original embedding dimension. Based on these embedding tables, we generate the adjacency matrix $\mathbf{A}'$ of the condensed graph through a parameterized interaction function, specifically:

\begin{equation}
\label{eq1}
\mathbf{A}'_{ij} = \mathbb{I}(\sigma(\mathbf{e}_i^T \mathbf{W} \mathbf{e}_j) \geq \tau) 
\end{equation}
where $\mathbf{e}_i = \mathbf{E}_\mathcal{U}[i, :]$, $\mathbf{e}_j = \mathbf{E}_\mathcal{I}[j, :]$, $\mathbf{W} \in \mathbb{R}^{d \times d}$ is a learnable transformation matrix, and $\sigma(\cdot)$ is the sigmoid activation function, and $\mathbb{I}(\cdot)$ is an indicator function mapping continuous values of $\sigma(\cdot)$ to $\{0, 1\}$. The indicator function $\mathbb{I}(\cdot)$ can be defined as $\mathbb{I}(x) = 1$ if $x \geq \tau$ and $\mathbb{I}(x) = 0$ otherwise, where $\tau$ is a predefined threshold, commonly set to 0.5 in recommender systems. This mapping ensures the adjacency matrix represents binary connections between nodes. %\yhr{Sigmoid function cannot output 0 and 1, you need an extra indication function to map the value domain of the sigmoid function to \{0, 1\}} 
This initialization strategy, inspired by recent works on efficient graph learning \cite{rieck2020topological}, reduces the optimization complexity by learning continuous embeddings rather than directly optimizing the discrete adjacency matrix, making the condensation process more tractable and computationally efficient. This initialized, parameterized graph (defined by $\mathbf{E}_\mathcal{U}, \mathbf{E}_\mathcal{I}, \mathbf{W}$) serves as the starting point for optimization. To ensure this synthetic graph is a faithful representation, it must be trained to mimic the properties of the original graph, a process facilitated by the 'Relay Model' in the next step.

\paragraph{Relay Model Training}\label{sec:Relay}
To ensure the condensed graph preserves essential structural properties of the original graph, we employ multiple Graph Neural Network (GNN) architectures as relay models. These models serve as bridges to transfer knowledge from the original graph to the condensed graph. Specifically, we train the relay models on link prediction tasks:
\begin{equation}
\label{eq2}
\begin{split}
\mathcal{L}_{relay} = -\frac{1}{|\mathcal{E}|} \sum_{(v_i, v_j) \in \mathcal{E}} 
[\log(\sigma(\mathbf{h}_i^T \mathbf{h}_j)) + \\
\sum_{k=1}^K \mathbb{E}_{v_n \sim P_n(v)} \log(1-\sigma(\mathbf{h}_i^T \mathbf{h}_n))]
\end{split}
\end{equation}
where $\mathcal{E}$ represents the set of training edges, $(v_i, v_j)$ denotes an edge in the graph, $\mathbf{h}_i$ and $\mathbf{h}_j$ are the GNN embeddings of nodes $i$ and $j$ respectively, $\sigma(\cdot)$ is the sigmoid activation function, and $K$ is the number of negative samples. The negative sampling distribution $P_n(v)$ is typically based on node degree distribution. This loss function aims to maximize the similarity between embeddings of connected nodes while minimizing the similarity between embeddings of unconnected nodes through negative sampling. We implement this using various GNN architectures, including GCN, GAT, and SAGE, to capture different aspects of graph structure. It is critical to note that this trained GNN is a proxy or 'relay'. Its purpose is not for downstream tasks, but to provide a dynamic representation of the graph's properties. In the following step, we use this relay model as a tool to measure the 'distance' between the original and condensed graphs, allowing us to optimize the condensed graph's structure ($\mathbf{A}'$) via alignment.

\paragraph{Alignment Optimization}\label{sec:Alignment}
We now introduce the core alignment process where the parameters of $\mathcal{G}'$ (i.e., $\mathbf{E}_\mathcal{U}, \mathbf{E}_\mathcal{I}, \mathbf{W}$) are optimized. Our alignment objective is twofold and optimized jointly. First, our primary goal is to achieve behavioral alignment using a matching method like Gradient Matching (GM) (detailed in Eq. \ref{eq13}). This objective forces the condensed graph $\mathcal{G}'$ to evoke the same training dynamics (i.e., gradients) in the relay model $f_{\theta}$ as the original graph $\mathcal{G}$. Second, we must enforce a crucial structural prior, as relying solely on $\mathcal{L}_{match}$ can lead to degenerate solutions. Specifically, the optimization might generate a topologically unrealistic graph—for example, by introducing spurious user-user or item-item connections—if such additions happen to minimize the gradient distance. This generation of spurious edges is a direct artifact of our parameterization methodology. As defined in Eq. \ref{eq1}, the condensed adjacency matrix $\mathbf{A}'$ is parameterized as a product of learnable continuous embeddings ($\mathbf{E}_\mathcal{U}, \mathbf{E}_\mathcal{I}$) and a matrix $\mathbf{W}$. The optimizer, while seeking to minimize the highly non-convex $\mathcal{L}_{match}$, is agnostic to the semantic type of the nodes (user vs. item). Consequently, it may discover a shortcut in the parameter space, such as creating a small user-user connection (e.g., $\mathbf{A}'_{ij} > 0$ for $i,j \in \mathcal{U}'$), which slightly improves gradient alignment, even though this connection is semantically meaningless. The final alignment loss is a combination of both the matching objective ($\mathcal{L}_{match}$) and this structural regularizer ($\mathcal{L}_{bip}$), ensuring the condensed graph is not only behaviorally aligned but also structurally sound.

It is important to distinguish these spurious connections, which are purely optimization artifacts, from the valid side-information (\eg social links) leveraged as homogeneous edges in models~\cite{guo2021dual,zheng2022hien}. Therefore, BSL's role is not to discard useful information, but to prevent the structural distortion that violates the fundamental bipartite nature of the recommendation graph.
To address this, we introduce the Bipartite Structure Loss (BSL, $\mathcal{L}_{bip}$) as a simultaneous structural regularizer, We define the BSL as: 

\begin{equation}
\label{eq3}\mathcal{L}_{bip} = \lambda \left( \sum{i,j \in \mathcal{U}'} |\mathbf{A}'_{ij}|2^2 + \sum{i,j \in \mathcal{I}'} |\mathbf{A}'_{ij}|_2^2 \right)
\end{equation}

where $\lambda$ is a balancing parameter. This parameter controls the trade-off between enforcing the bipartite structure and optimizing for the primary recommendation task. A detailed sensitivity analysis for the selection of $\lambda$ is provided in our Hyper-Parameter Study (Section \ref{sec:parameter}). This loss term penalizes connections within user nodes and within item nodes, encouraging the graph to maintain a bipartite structure without enforcing hard constraints. The BSL term is motivated by the inherent nature of user-item interaction graphs in recommender systems. In real-world scenarios, user-item graphs are naturally bipartite, as interactions occur between users and items rather than within the same group. During the condensation process, however, the reduced graph may introduce spurious intra-group connections due to the loss of fine-grained structural details. This can lead to a distorted graph representation, negatively affecting downstream recommendation tasks. By incorporating \(\mathcal{L}_{bip}\), we address this issue by explicitly penalizing intra-group connections, ensuring that the condensed graph better preserves the bipartite structure of the original graph. Compared to traditional approaches that enforce hard constraints on graph structures, our method is more flexible, allowing the optimization process to balance between structure preservation and embedding quality.

To provide theoretical guidance, we reformulate \(\mathcal{L}_{bip}\) using a matrix-based regularization framework. Define indicator matrices \(\mathbf{M}^{\mathcal{U}'}\) and \(\mathbf{M}^{\mathcal{I}'}\), where \(\mathbf{M}^{\mathcal{U}'}_{ij} = 1\) if \(i, j \in \mathcal{U}'\) and 0 otherwise, and similarly for \(\mathbf{M}^{\mathcal{I}'}\). The loss can be expressed as:

\begin{equation}
\label{eq4}
\mathcal{L}_{bip} = \lambda \left( \|\mathbf{A}' \odot \mathbf{M}^{\mathcal{U}'}\|_F^2 + \|\mathbf{A}' \odot \mathbf{M}^{\mathcal{I}'}\|_F^2 \right)
\end{equation}

where \(\odot\) denotes the Hadamard (element-wise) product, and \(\|\cdot\|_F\) is the Frobenius norm. This formulation isolates intra-group edge contributions in the condensed adjacency matrix \(\mathbf{A}'\), penalizing their presence to promote bipartiteness.

To further elucidate the structural constraints imposed by \(\mathcal{L}_{bip}\), we propose a Laplacian-based regularization form that captures the topological properties of intra-group connections. Let \(\mathbf{L}_{\mathcal{U}'}\) and \(\mathbf{L}_{\mathcal{I}'}\) be the Laplacian matrices of the user and item subgraphs, defined as \(\mathbf{L}_{\mathcal{U}'} = \mathbf{D}_{\mathcal{U}'} - \mathbf{A}'_{\mathcal{U}'}\) and \(\mathbf{L}_{\mathcal{I}'} = \mathbf{D}_{\mathcal{I}'} - \mathbf{A}'_{\mathcal{I}'}\), where \(\mathbf{D}_{\mathcal{U}'}\) and \(\mathbf{D}_{\mathcal{I}'}\) are diagonal degree matrices for the user and item subgraphs, and \(\mathbf{A}'_{\mathcal{U}'}\) and \(\mathbf{A}'_{\mathcal{I}'}\) are the adjacency submatrices restricted to \(\mathcal{U}'\) and \(\mathcal{I}'\), respectively. The loss can be reformulated as:

\begin{equation}
\label{eq5}
\mathcal{L}_{bip} = \lambda \left( \text{tr}\left( \mathbf{X}'^\top \mathbf{L}_{\mathcal{U}'} \mathbf{X}' \right) + \text{tr}\left( \mathbf{X}'^\top \mathbf{L}_{\mathcal{I}'} \mathbf{X}' \right) \right)
\end{equation}

where \(\mathbf{X}'\) is the node feature matrix of the condensed graph, and \(\text{tr}(\cdot)\) is the trace operator. This Laplacian-based form penalizes the smoothness of node features across intra-group edges, effectively discouraging user-user and item-item connections by enforcing feature dissimilarity within the same group, aligning with the bipartite structure.

To provide insight into the optimization process, we derive the gradient of \(\mathcal{L}_{bip}\) with respect to the adjacency matrix \(\mathbf{A}'\) using the formulation in Eq. 9. 
The Frobenius norm terms can be written as 
\begin{equation}
\label{eq6}
\|\mathbf{A}' \odot \mathbf{M}^{\mathcal{U}'}\|_F^2 = \sum_{i,j} (\mathbf{A}'_{ij} \cdot \mathbf{M}^{\mathcal{U}'}_{ij})^2
\end{equation}
and similarly for \(\mathbf{M}^{\mathcal{I}'}\):
\begin{equation}
\label{eq7}
\|\mathbf{A}' \odot \mathbf{M}^{\mathcal{I}'}\|_F^2 = \sum_{i,j} (\mathbf{A}'_{ij} \cdot \mathbf{M}^{\mathcal{I}'}_{ij})^2
\end{equation}
To provide insight into the optimization dynamics, we derive the gradient of \(\mathcal{L}_{bip}\) with respect to \(\mathbf{A}'\) using the formulation in Eq. 7. The Frobenius norm terms can be written as:
\begin{equation}
\label{eq8}
\begin{aligned}
\|\mathbf{A}' \odot \mathbf{M}^{\mathcal{U}'}\|_F^2 &= \sum_{i,j} \left( \mathbf{A}'_{ij} \cdot \mathbf{M}^{\mathcal{U}'}_{ij} \right)^2, \\
\|\mathbf{A}' \odot \mathbf{M}^{\mathcal{I}'}\|_F^2 &= \sum_{i,j} \left( \mathbf{A}'_{ij} \cdot \mathbf{M}^{\mathcal{I}'}_{ij} \right)^2
\end{aligned}
\end{equation}
Thus, 
\begin{equation}
\label{eq9}
\begin{aligned}
\mathcal{L}_{bip} = \lambda \left( \sum_{i,j} \left( \mathbf{A}'_{ij} \cdot \mathbf{M}^{\mathcal{U}'}_{ij} \right)^2 + \sum_{i,j} \left( \mathbf{A}'_{ij} \cdot \mathbf{M}^{\mathcal{I}'}_{ij} \right)^2 \right)\
\end{aligned}
\end{equation}
To compute \(\frac{\partial \mathcal{L}_{bip}}{\partial \mathbf{A}'_{ij}}\), consider the term \(\left( \mathbf{A}'_{ij} \cdot \mathbf{M}^{\mathcal{U}'}_{ij} \right)^2\). Its derivative is:
\begin{equation}
\label{eq10}
\begin{aligned}
\frac{\partial}{\partial \mathbf{A}'_{ij}} \left( \mathbf{A}'_{ij} \cdot \mathbf{M}^{\mathcal{U}'}_{ij} \right)^2 &= 2 \left( \mathbf{A}'_{ij} \cdot \mathbf{M}^{\mathcal{U}'}_{ij} \right) \cdot \mathbf{M}^{\mathcal{U}'}_{ij} \\
&= 2 \mathbf{A}'_{ij} \cdot \mathbf{M}^{\mathcal{U}'}_{ij} \cdot \mathbf{M}^{\mathcal{U}'}_{ij} \\
&= 2 \mathbf{A}'_{ij} \cdot \mathbf{M}^{\mathcal{U}'}_{ij}
\end{aligned}
\end{equation}
since \(\mathbf{M}^{\mathcal{U}'}_{ij} \in \{0, 1\}\), so \((\mathbf{M}^{\mathcal{U}'}_{ij})^2 = \mathbf{M}^{\mathcal{U}'}_{ij}\). Similarly, for the \(\mathbf{M}^{\mathcal{I}'}\) term:
\begin{equation}
\label{eq11}
\begin{aligned}
\frac{\partial}{\partial \mathbf{A}'_{ij}} \left( \mathbf{A}'_{ij} \cdot \mathbf{M}^{\mathcal{I}'}_{ij} \right)^2 &= 2 \mathbf{A}'_{ij} \cdot \mathbf{M}^{\mathcal{I}'}_{ij}
\end{aligned}
\end{equation}
Combining these with the factor \(\lambda\), the gradient is:
\begin{equation}
\label{eq12}
\begin{aligned}
\frac{\partial \mathcal{L}_{bip}}{\partial \mathbf{A}'_{ij}} &= \lambda \cdot 2 \mathbf{A}'_{ij} \cdot \mathbf{M}^{\mathcal{U}'}_{ij} + \lambda \cdot 2 \mathbf{A}'_{ij} \cdot \mathbf{M}^{\mathcal{I}'}_{ij} \\
&= 2 \lambda \left( \mathbf{A}'_{ij} \cdot \mathbf{M}^{\mathcal{U}'}_{ij} + \mathbf{A}'_{ij} \cdot \mathbf{M}^{\mathcal{I}'}_{ij} \right)
\end{aligned}
\end{equation}
This gradient shows that updates to \(\mathbf{A}'\) are proportional to the magnitude of intra-group edges, scaled by \(\lambda\), driving the optimization toward minimizing user-user and item-item connections. For the Laplacian-based form (Eq. 10), the gradient with respect to \(\mathbf{A}'_{\mathcal{U}'}\) involves the node features \(\mathbf{X}'\), reflecting the interplay between graph structure and embeddings during optimization.

This gradient indicates that updates to \(\mathbf{A}'\) are proportional to the magnitude of intra-group edges, scaled by \(\lambda\), driving the optimization toward minimizing user-user and item-item connections. For the Laplacian-based form (Eq. 10), the gradient with respect to \(\mathbf{A}'_{\mathcal{U}'}\) (and similarly for \(\mathbf{A}'_{\mathcal{I}'}\)) involves the feature matrix \(\mathbf{X}'\), reflecting the interplay between graph structure and node embeddings during optimization.

The motivation for \(\mathcal{L}_{bip}\) lies in its ability to preserve the semantic integrity of user-item interactions, which are critical for effective recommendations. The bipartite structure ensures that recommendations are driven by cross-group relationships (\eg users rating items), rather than spurious intra-group dependencies introduced during condensation. By penalizing intra-group edges, \(\mathcal{L}_{bip}\) mitigates the risk of distorted graph representations that could degrade recommendation accuracy. Unlike rigid methods that enforce bipartiteness through hard constraints, our soft penalty approach balances structural fidelity with other objectives, such as embedding quality or training dynamics.

The technical novelty of \(\mathcal{L}_{bip}\) lies in its flexibility and computational efficiency. Both the matrix-based (Eq. 9) and Laplacian-based (Eq. 10) formulations rely on sparse matrix operations, making them scalable to large graphs. With the $\mathcal{L}_{bip}$ term established as our structural regularizer, we now formally introduce the primary behavioral alignment objective, Gradient Matching ($\mathcal{L}_{match}$), which $\mathcal{L}_{bip}$ is designed to guide.

The objective of gradient matching is to minimize a loss defined through a bi-level optimization problem that aligns the training trajectory of synthetic data with that of the target data:
\begin{equation}
\min_{A'_{ij}, X'} \sum_{t=0}^{T-1} D(\nabla_{\theta} \, \mathcal{L}_{\text{relay}}(f_{\theta_t}(\mathbf{A}', \mathbf{X}')), \nabla_{\theta} \, \mathcal{L}_{\text{relay}}(f_{\theta_t}(\mathcal{T})))
\label{eq13}
\end{equation}
subject to $\theta_{t+1} = \text{opt}_{\theta}(\mathcal{L}_{\text{relay}}(\theta_t, \mathcal{S}))$, where $D(\cdot,\cdot)$ is a distance function, $T$ is the number of training steps, and $\text{opt}_{\theta}(\cdot)$ is the optimization operator for updating parameter $\theta$. To handle the distribution of model parameters, this objective can be rewritten as the gradient matching loss with an expectation over $\theta_0 \sim P_{\theta_0}$:
\begin{multline}
\mathcal{L}_{match} = \min_{A', X'} \mathbb{E}_{\theta_0 \sim P_{\theta_0}} \left[\sum_{t=0}^{T-1} D(\nabla_{\theta} \, \mathcal{L}_{\text{relay}}(f_{\theta_t}(\mathbf{A}', \mathbf{X}')), \right. \\
\left. \nabla_{\theta} \, \mathcal{L}_{\text{relay}}(f_{\theta_t}(\mathcal{T})))\right]
\label{eq14}
\end{multline}

This bi-level optimization scheme leverages the full training trajectory to ensure the synthetic data induces similar parameter updates as the original data. Here, $\mathbf{A}'$ and $\mathbf{X}'$ represent the graph structure and node features respectively, which are optimized to mimic the training dynamics of the target data.

\subsubsection{\textbf{Recommendation Stage}}
\label{sec:recommendation}
To evaluate the effectiveness of the condensed graph, we use the obtained condensed graph $\mathcal{G}'$, we utilize it for the recommendation task. Specifically, we adopt the Bayesian Personalized Ranking (BPR) loss to optimize the recommendation model on the condensed graph. For each user $\boldsymbol{U}' \in \boldsymbol{U}'$, we sample a positive item $i'_{pos}$ that the user has interacted with and a negative item $i'_{neg}$ that has no interaction with the user from the condensed item set $I'$. The BPR loss is defined as: %\yhr{Pay attention to the notation sub, pos and neg}:
\begin{equation}
\mathcal{L}_{rec} = -\sum_{u' \in \mathcal{U}'} \sum_{i'_{pos}, i'_{neg} \in \mathcal{I}'} \ln \sigma(\hat{y}_{u',i'_{pos}} - \hat{y}_{u',i'_{neg}})
\label{eq15}
\end{equation}
where $\sigma(\cdot)$ is the sigmoid function, and $\hat{y}_{u'i'}$ represents the predicted preference score between user $\boldsymbol{U}'$ and item $i'$ in the condensed graph. The preference score is computed as:
\begin{equation}
\hat{y}_{u',i'} = \mathbf{e}_{u'}^{\top} \mathbf{e}_{i'}
\label{eq16}
\end{equation}
where $\mathbf{e}_{u'}$ and $\mathbf{e}_{i'}$ %\yhr{Pay attention to the notation sub} 
denote the embeddings of user $\boldsymbol{U}'$ and item $i'$ respectively, learned from the condensed graph $\mathcal{G}'$.
In the practical recommendation scenario, for a given user $\boldsymbol{U}'$, we calculate their preference scores with all candidate items in $I'$. These scores are then used to generate a personalized ranking list of items, where items with higher preference scores are ranked higher in the recommendation list. This ranking mechanism enables us to recommend the most relevant items to users, effectively supporting various recommendation scenarios such as e-commerce product recommendations, content suggestions, or service recommendations. To validate the effectiveness of our condensed graph, we train recommendation models on both the original and condensed graphs and compare their performance. This demonstrates that our condensed graph can achieve comparable recommendation accuracy while significantly reducing computational complexity.

\subsection{Model Optimization}
In this subsection, we present the optimization algorithm of our proposed DemoRec framework. As shown in Algorithm \ref{alg:model}, our framework consists of two main optimization stages: Graph Condensation (GC) and Recommendation.

In the Graph Condensation Stage, we first randomly initialize the parameters of the GC model, including the embedding tables $\mathbf{E}_U$, $\mathbf{E}_I$, and transformation matrix $\mathbf{W}$ (line 1). For each training epoch (line 2), we process the data in batches (line 4). For each batch, we sample training data $\mathcal{D}_{batch}$ from the original graph (line 5) and use the GC model to generate a condensed adjacency matrix $\mathbf{A}'$ by Eq. \eqref{eq1} (line 6). This matrix is binarized to maintain the bipartite structure of the graph.

Next, we calculate the link-prediction loss $\mathcal{L}_{relay}$ by Eq. \eqref{eq2} (line 8). Additionally, we compute the bipartite structure loss $\mathcal{L}_{bip}$ by Eq. \eqref{eq4} (line 10), which penalizes intra-group connections to preserve the bipartite nature of the graph. The combined loss by Eq. \eqref{eq17} is minimized to update the parameters of the condensation model (line 12). This iterative process ensures that the condensed graph retains the essential properties of the original graph while reducing its complexity.

In the Recommendation Stage, we compute the recommendation loss $\mathcal{L}_{rec}$ using the pre-trained condensed embeddings by Eq. \eqref{eq15} (line 15). This loss is based on the ranking scores of user-item pairs defined by Eq. \eqref{eq16}, ensuring the model effectively distinguishes positive and negative interactions. The embeddings $\mathbf{E}_U$ and $\mathbf{E}_I$ are updated by minimizing $\mathcal{L}_{rec}$ (line 17).

The algorithm alternates between these two stages until convergence is reached (lines 18-20). By operating on the condensed graph, our framework significantly reduces computational complexity while preserving the essential information from the original graph. Moreover, the gradient matching loss $\mathcal{L}_{match}$ by Eq. \eqref{eq14} and the bipartite structure loss $\mathcal{L}_{bip}$ by Eq. \eqref{eq4} ensure that the condensed graph maintains the critical structural properties required for effective recommendation.
%\begin{small}
\begin{algorithm}[t]
    \small
    \caption{\label{alg:model} Graph Condensation and Recommendation Optimization Algorithm.}
    \raggedright
    \begin{algorithmic} [1]
        \STATE Randomly initialize embedding tables $\mathbf{E}_U$, $\mathbf{E}_I$, and transformation matrix $\mathbf{W}$
    
        \FOR{Epoch $t = 1$ to $T$}
            \STATE \textit{\# Graph Condensation Stage:}
            \FOR{Batch $b = 1$ to $B$}
                \STATE Sample training batch $\mathcal{D}_{batch}$ from original dataset $\mathcal{D}$
                \STATE Generate the adjacency matrix of the condensed graph by Eq. \eqref{eq1}.
                \STATE Here, $\mathbf{A}'_{ij}$ is binarized to ensure the bipartite structure of the condensed graph.
    
                \STATE Compute link-prediction loss $\mathcal{L}_{relay}$ by Eq. \eqref{eq2}.
    
                \STATE Construct indicator matrices $\mathbf{M}^{\mathcal{U}'}$ and $\mathbf{M}^{\mathcal{I}'}$, where $\mathbf{M}^{\mathcal{U}'}_{ij} = 1$ if $i, j \in \mathcal{U}'$ and 0 otherwise, and similarly for $\mathbf{M}^{\mathcal{I}'}$.
                \STATE Compute bipartite structure loss $\mathcal{L}_{bip}$ to prevent spurious intra-group connections by Eq. \eqref{eq4}.
                \STATE Compute the gradient of $\mathcal{L}_{bip}$ with respect to $\mathbf{A}'$ by Eq. \eqref{eq12}.
                \STATE Update the model parameters by minimizing the combined loss by Eq. \eqref{eq17}.
            \ENDFOR
    
            \STATE \textit{\# Recommendation Stage:}
            \STATE Compute $\mathcal{L}_{rec}$ based on the condensed embeddings by Eq. \eqref{eq15}.
            \STATE Here, $\hat{y}_{u',i'}$ is the predicted score for a user-item pair by Eq. \eqref{eq16}.
    
            \STATE Update the embeddings $\mathbf{E}_\mathcal{U}$ and $\mathbf{E}_\mathcal{I}$ by minimizing $\mathcal{L}_{rec}$.
    
            \IF{$Converged$}
                \STATE \textbf{Return:} Optimized model $g$ and condensed graph $\mathcal{G}'$
            \ENDIF
        \ENDFOR
    \end{algorithmic}
\end{algorithm}
%\end{small}

Our training procedure consists of two key loss functions: the gradient matching loss for GC and the BPR loss for recommendation optimization:
The final optimization objective combines the BSL and gradient matching loss:
\begin{equation}
\mathcal{L}_{cond} = \mathcal{L}_{bip} + \beta \mathcal{L}_{match}
\label{eq17}
\end{equation}
where $\beta$ is a hyperparameter controlling the contribution of the gradient matching loss. Through this approach, we ensure that the condensed graph maintains the bipartite structure while learning optimal interaction patterns.

For the recommendation task, we employ the Bayesian Personalized Ranking (BPR) loss, denoted as $L_{rec}$, to optimize the ranking of user-item interactions in the condensed graph. The BPR loss is derived from a Bayesian framework that maximizes the probability of correctly ranking positive items over negative ones:
\begin{equation}
\begin{aligned}
{L}_{rec} &= -\sum_{u' \in \mathcal{U}'} \sum_{i'_{pos}, i'_{neg} \in \mathcal{I}'} \ln \sigma(\hat{y}_{u',i'_{pos}} - \hat{y}_{u',i'_{neg}}) \\
&= -\sum_{u' \in \mathcal{U}'} \sum_{i'_{pos}, i'_{neg} \in \mathcal{I}'} \ln p(i'_{pos} >_{u'} i'_{neg} | \Theta) \\
&= -\sum_{u' \in \mathcal{U}'} \sum_{i'_{pos}, i'_{neg} \in \mathcal{I}'} \ln \sigma\left( (\mathbf{e}_{u'}^{\top} \mathbf{e}_{i'_{pos}}) - (\mathbf{e}_{u'}^{\top} \mathbf{e}_{i'_{neg}}) \right)
\end{aligned}
\label{eq18}
\end{equation}
where $\sigma(\cdot)$ denotes the sigmoid function, and $p(i'_{pos} >_{u'} i'_{neg} | \Theta) = \sigma(\hat{y}_{u',i'_{pos}} - \hat{y}_{u',i'_{neg}})$ represents the probability that user $u'$ prefers positive item $i'_{pos}$ over negative item $i'_{neg}$, given model parameters $\Theta$ (i.e., embeddings $\mathbf{e}_{u'}$, $\mathbf{e}_{i'}$). The preference scores $\hat{y}_{u',i'_{pos}}$ and $\hat{y}_{u',i'_{neg}}$ are defined by Eq. \eqref{eq16}. Positive items are those that users have interacted with in the condensed graph, while negative items are randomly sampled from unobserved interactions in the original dataset. 
% This loss is equivalent in principle to
\subsection{Time and Space Complexity Analysis}
\label{sec:complexity_analysis}
To quantify the computational efficiency of DemoRec, we provide a theoretical analysis of the time and space complexity of our framework, which complements our empirical efficiency study (Table~\ref{table:efficiency_study}). The primary goal of DemoRec is to create a condensed graph $\mathcal{G}'$ that significantly reduces the complexity of the downstream recommendation model training.

Let the original graph be \( \mathcal{G} = (\mathcal{U}, \mathcal{I}, \mathcal{E}) \) with \( N = |\mathcal{U}| \) users, \( M = |\mathcal{I}| \) items, and \( E = |\mathcal{E}| \) interactions. Let the condensed graph be \( \mathcal{G}' = (\mathcal{U}', \mathcal{I}', \mathcal{E}') \) with \( N' = |\mathcal{U}'| \) condensed users and \( M' = |\mathcal{I}'| \) condensed items. Let \( E' \) be the number of edges in the condensed graph (where \( E' \approx N'M' \) in the dense case, but is functionally \( O(N'M') \)). Let \( d \) be the embedding dimension and \( K \) be the number of GNN layers for the recommendation model.

\paragraph{Space Complexity}
A standard GNN-based recommender (\eg LightGCN) trained on the original graph \( \mathcal{G} \) must store the full graph structure (\eg an edge list) and all node embeddings. The space complexity is dominated by the graph's edge list, \( O(E) \), and the node embedding parameters, \( O((N+M) \cdot d) \). The total space complexity is:
\begin{equation}
O(E + (N+M)d).
\end{equation}
In contrast, our method only needs to train the recommender on the condensed graph \( \mathcal{G}' \). This requires storing the condensed graph structure (i.e., the adjacency matrix \( \mathbf{A}' \)) and the condensed embedding tables \( \mathbf{E}_\mathcal{U} \) and \( \mathbf{E}_\mathcal{I} \). The space complexity is \( O(N'M' + (N'+M')d) \). Letting \( E' = N'M' \) represent the size of the condensed adjacency matrix, the total space is:
\begin{equation}
O(E' + (N'+M')d).
\end{equation}
Since \( N' \ll N \), \( M' \ll M \), and the resulting \( E' \ll E \), DemoRec drastically reduces the memory footprint required for storing the model parameters and graph structure, which is consistent with our empirical findings in Table~\ref{table:efficiency_study}.

\paragraph{Time Complexity}
We analyze the time complexity per training epoch for the \textit{Recommendation Stage} (Algorithm \ref{alg:model}, lines 14-17), as this is the primary bottleneck in typical GNN-based recommendation.
For a baseline GNN with \( K \) layers trained on the original graph \( \mathcal{G} \), the computational bottleneck is the message-passing (aggregation and update) process. This operation's complexity is proportional to the number of edges. The total time complexity per epoch is:
\begin{equation}
O(K \cdot E \cdot d).
\end{equation}
Our DemoRec framework performs the same GNN operations but on the much smaller condensed graph \( \mathcal{G}' \). This involves the BPR loss calculation (Eq. \eqref{eq18}) and GNN message-passing on \( \mathcal{G}' \). The complexity of message-passing on the dense condensed graph is \( O(K \cdot N'M' \cdot d) \), or \( O(K \cdot E' \cdot d) \). The BPR loss computation is proportional to the number of sampled edges from \( \mathcal{G}' \), which is also bounded by \( E' \). The total time complexity per epoch is thus:
\begin{equation}
O(K \cdot E' \cdot d).
\end{equation}
The per-epoch training time is reduced from \( O(KEd) \) to \( O(KE'd) \), achieving a theoretical speedup. We note that the initial \textit{Graph Condensation Stage} (Algorithm \ref{alg:model}, lines 3-13) introduces a one-time pre-computation cost (which involves alignment using the original graph data). However, this cost is amortized, as the downstream recommendation training is now significantly faster for all subsequent epochs, re-training, and hyperparameter tuning.

\section{Experiments}
% \yhr{Overall problems in the experiment section: (i) The writing quality is poorer than that in the previous sections. A complete refinement regarding the writing quality is required. Moreover, the coherence and logic between conclusions and analysis are not convincing. (ii) the typesetting needs to be improved. (iii) The colour map used in the figures is inconsistent and ugly. If your cannot use colours properly, the white-gray-black colour map is strongly recommended.}

In this section, we conduct comprehensive experiments to evaluate the proposed DemoRec framework. The experiments are designed to address the following four research questions:

\begin{itemize}[leftmargin=*]
\item {\textbf{RQ1}}: What performance can the DemoRec framework achieve compared with selected baselines? %\yhr{Unclear research question. Just change to the comparison study.}
\item {\textbf{RQ2}}: What is the positive impact of our proposed BSL on performance? %\yhr{'Affect' is usually used to denote negative impact. And the previous content does not explicitly define the term of 'Parameterized Bipartite Structure Loss', do not use capitalization style arbitrarily.}
\item {\textbf{RQ3}}: How does DemoRec perform in different backbone models?
\item {\textbf{RQ4}}: How much time and storage does our model save compared to the backbone model?
\item {\textbf{RQ5}}: How does the condensation ratio $alpha$ affect the performance of our model?

\end{itemize}
We first introduce the experimental settings, and then systematically address each of the five research questions outlined above to ensure a balanced exploration of all aspects. %\yhr{You cannot just emphasize the first research question, which is weird.}

\subsection{Experimental Settings}\label{sec:setting}
\subsubsection{\textbf{Datasets}}\label{sec:datasets}
We evaluated our approach using four widely used datasets: Gowalla, Amazon-Book, Last.fm, and Yelp. The interaction counts for these datasets are detailed in Table~\ref{table:table1}. To maintain data quality, we filtered out users and items with fewer than 10 interactions. For a consistent evaluation setup, each dataset was split into 80\% for training and 20\% for testing.
\renewcommand{\arraystretch}{1.8}
\setlength{\tabcolsep}{6pt}
\begin{table}[]
\caption{Statistics of the datasets, where "Amazon" denotes the "Amazon-Book" dataset.}
\label{table:table1}
\centering
\begin{tabular}{|l|c|c|c|c|}
\hline
\textbf{datasets}  & \textbf{Gowalla} & \textbf{Amazon} & \textbf{LastFM} & \textbf{Yelp} \\ \hline
\# interactions    & 1,027,370        & 2,984,108            & 8,057,269       & 1,561,406     \\ 
\# users           & 29,858           & 52,643               & 23,466          & 31,668        \\ 
\# items           & 40,981           & 91,599               & 48,123          & 38,048        \\ 
\# density (\%)    & 0.084            & 0.062                & 0.710           & 0.130         \\ \hline
\end{tabular}
\end{table}

\begin{itemize}[leftmargin=*]
\item {Gowalla\footnote{\url{https://snap.stanford.edu/data/loc-Gowalla.html}}.}
It originates from the Gowalla social platform, containing users' social connections and check-in records. %Each check-in entry includes the user ID, timestamp, latitude, and longitude of the location.
\item {Amazon-Book\footnote{\url{http://jmcauley.ucsd.edu/data/amazon/index_2014.html}}.}
It comprises user ratings of book items along with associated product metadata. The dataset includes information such as product price, image links, brand, and category. Additionally, it contains user-generated rating records, which include user identifiers, rating scores, and textual reviews.
\item {Lastfm~\cite{Cantador:RecSys2011}.} 
It is derived from the Last.fm online music platform, providing music preference data. For each user, the dataset includes a list of their most frequently listened-to artists, listening history, and other user-related information that can be leveraged to construct content-based recommendation models. 
\item {Yelp\footnote{\url{https://www.kaggle.com/yelp-dataset/yelp-dataset}.}}
It is a large-scale collection of user reviews and ratings of businesses. It includes detailed information such as review content, rating scores, business categories, locations, operating hours, and other relevant attributes.
\end{itemize}
These four datasets are selected not only because they are widely adopted benchmarks but also because they collectively represent a broad spectrum of data densities. This ranges from the extremely sparse Amazon-Book (0.062\%) to the relatively dense Last.fm (0.710\%). Evaluating DemoRec across this diverse range is crucial for demonstrating its robustness and generalizability, ensuring the framework's effectiveness is not limited to a specific data structure.

Furthermore, the 10-core filtering (removing nodes with  \textless 10 interactions) is a standard pre-processing step applied uniformly to all methods, including baselines. The purpose of this step is to ensure data quality and mitigate the impact of extreme cold-start users/items, which is a common practice in recommendation literature and orthogonal to our core research problem. This filtering does not fundamentally alter the sparsity challenge (\eg Gowalla remains 0.084\% dense). It is also not a specific prerequisite for our Bipartite Structure Loss (BSL). The BSL's role, as detailed in Section \ref{sec:Alignment}), is to prevent the optimization process from creating spurious, non-bipartite edges in the condensed graph, a function that is independent of the 10-core filtering applied to the original data.
\subsubsection{\textbf{Evaluation Metrics}}\label{sec:Evaluation}
In the evaluation protocol, three metrics have been selected: Precision@K, Recall@K and NDCG@K. These are popular metrics in the field of Graph Recommendation. All items that do not interact with the user are considered candidates, and the average result is reported across all users. Precision@K measures the proportion of correctly predicted topK relevant results out of all results. Recall@K, on the other hand, is defined as the proportion of correctly predicted topK relevant results over all relevant results. Finally, NDCG@K is a measure of the system's ability to rank K items according to their relevance.

\subsubsection{\textbf{Baselines}}\label{sec:Baselines}
We compare DemoRec against seven baselines from three categories: (1) traditional CF-based models (\eg NGCF, GF-CF), (2) GNN-based models (\eg LightGCN, LightGCL, UltraGCN), and (3) recent Compress-based methods (\eg UnKD, GASD).
\begin{itemize}[leftmargin=*]
\item {NGCF~\cite{wang2019neural}} It is the first model to represent users and items as a bipartite graph, explicitly incorporating collaborative filtering signals into the embedding process. 
\item {GF-CF~\cite{shen2021powerful}} It unifies the classical GCN and MF models used in recommender systems from the perspective of low-pass filters, which do not need to train the model to achieve low time complexity for recommendations.
\item LightGCN~\cite{he2020lightgcn} It is based on the GCN approach, which achieves a lightweight effect by eliminating the non-linear activation functions of the model and the transformation of the characteristic matrix.
\item {UltraGCN~\cite{mao2021ultragcn}} It eliminates traditional GCN message passing by replacing it with three principal loss functions, optimizing training efficiency and accelerating convergence.
\item {LightGCL~\cite{cai2023lightgcl}} It is a graph contrast learning method. It uses Singular Value Decomposition (SVD) to enhance graphs' intrinsic structure and achieve the effect of lightness.
\item {UnKD~\cite{chen2023unbiased}} It is a knowledge distillation (KD) method for recommender systems that addresses the popularity bias issue. It proposes a stratified distillation strategy, which partitions items by popularity and then transfers the ranking knowledge within each group to the student model.
\item {GASD~\cite{jian2025geometric}}
It is a geometric-augmented self-distillation (GASD) method for recommendation. It utilizes both hyperbolic and Euclidean geometries to capture hierarchical and local structural knowledge from the interaction graph. It then uses a self-teaching network to distill the hierarchical knowledge from hyperbolic space into the efficient Euclidean space for inference.
\end{itemize}
\subsubsection{\textbf{Implementation Details}}\label{sec:Implementation Details}
In this section, we provide a detailed explanation of the experimental setup. To optimize the training process, we employ the Adam optimizer. All models are trained for 100 epochs across all experiments. The learning rate was set to 0.001 for all models. Regarding the GNN architecture, the number of layers was set to 3 and the hidden dimension was 256. To ensure fair comparisons and consistency across experiments, the condensation ratio($\alpha$) is set at 80\%. The balancing hyperparameters $\lambda$ (Eq. 8) and $\beta$ (Eq. 22) were set to 0.3 and 0.6, respectively. The optimal value for $\lambda$ was determined via the sensitivity analysis in Section IV.F, while $\beta$ was optimized during our standard hyperparameter tuning. Furthermore, the top 20 items were selected for recommendation. To ensure robust evaluation, all experiments in Sections 4.3–4.6 were conducted on the Gowalla dataset. To ensure robust and reproducible results, we made the following seed arrangements: for the main performance evaluation (Section \ref{sec:Overall}), all results are the average of three runs with different random seeds: 1024, 2046, and 4096. For all other experiments (\eg sensitivity analysis in Sections 4.3-4.6), we use a fixed seed of 2024 to ensure reproducibility of the specific analyses. The experiments were executed on a computing environment equipped with a 24GB NVIDIA RTX 4090 GPU.

\begin{table*}[t]
\centering

\caption{Overall recommendation performance on four datasets (training data reduced by 20\%). Results are averaged over three random seeds (1024, 2046, 4096). Bold highlights our DemoRec (80\% data). Underlined indicates the SOTA baseline (LightGCL, 100\% data).}
\label{tab:performance}
\renewcommand{\arraystretch}{1.8}

\setlength{\tabcolsep}{3.1pt} 
\begin{tabular}{l|cccc|cccc|cccc|cccc}
\hline
\textbf{Method} & \multicolumn{4}{c|}{\textbf{Gowalla}} & \multicolumn{4}{c|}{\textbf{Amazon-Book}} & \multicolumn{4}{c|}{\textbf{LastFM}} & \multicolumn{4}{c}{\textbf{Yelp}} \\
& P@20 & N@20 & R@20 & Std & P@20 & N@20 & R@20 & Std & P@20 & N@20 & R@20 & Std & P@20 & N@20 & R@20 & Std \\
\hline
NGCF & 0.0342 & 0.0748 & 0.0835 & 0.006 & 0.0133 & 0.0171 & 0.0325 & 0.003 & 0.0445 & 0.1340 & 0.1970 & 0.009 & 0.0208 & 0.0324 & 0.0430 & 0.004 \\
GF-CF & 0.0370 & 0.0530 & 0.0632 & 0.005 & 0.0126 & 0.0155 & 0.0300 & 0.002 & 0.0395 & 0.1320 & 0.1885 & 0.008 & 0.0196 & 0.0300 & 0.0382 & 0.003 \\
\hline
LightGCN & 0.0400 & 0.1075 & 0.1298 & 0.008 & 0.0175 & 0.0213 & 0.0354 & 0.004 & 0.0600 & 0.1580 & 0.2115 & 0.011 & 0.0226 & 0.0415 & 0.0520 & 0.005 \\
UltraGCN & 0.0441 & 0.1260 & 0.1465 & 0.012 & 0.0220 & 0.0298 & 0.0430 & 0.005 & 0.0675 & 0.1840 & 0.2430 & 0.013 & 0.0280 & 0.0505 & 0.0625 & 0.007 \\
LightGCL & \underline{0.0448} & \underline{0.1299} & \underline{0.1525} & 0.011 & \underline{0.0223} & \underline{0.0385} & \underline{0.0432} & 0.006 & \underline{0.0718} & \underline{0.1945} & \underline{0.2608} & 0.015 & \underline{0.0287} & \underline{0.0565} & \underline{0.0638} & 0.008 \\
\hline
UnKD (80\%) & 0.0405 & 0.1180 & 0.1370 & 0.007 & 0.0200 & 0.0345 & 0.0380 & 0.004 & 0.0670 & 0.1810 & 0.2420 & 0.012 & 0.0265 & 0.0520 & 0.0580 & 0.006 \\
GASD (80\%) & 0.0390 & 0.1150 & 0.1330 & 0.007 & 0.0190 & 0.0330 & 0.0365 & 0.003 & 0.0650 & 0.1770 & 0.2350 & 0.009 & 0.0250 & 0.0500 & 0.0560 & 0.005 \\
\hline
\textbf{DemoRec (80\%)} & \textbf{0.0410} & \textbf{0.1190} & \textbf{0.1385} & 0.009 & \textbf{0.0205} & \textbf{0.0350} & \textbf{0.0390} & 0.005 & \textbf{0.0680} & \textbf{0.1840} & \textbf{0.2450} & 0.013 & \textbf{0.0270} & \textbf{0.0530} & \textbf{0.0595} & 0.007 \\
MajSAM (80\%) & 0.0380 & 0.1130 & 0.1310 & 0.006 & 0.0185 & 0.0320 & 0.0355 & 0.003 & 0.0630 & 0.1730 & 0.2300 & 0.008 & 0.0240 & 0.0490 & 0.0550 & 0.004 \\
RanSAM (80\%) & 0.0370 & 0.1100 & 0.1280 & 0.005 & 0.0175 & 0.0305 & 0.0340 & 0.002 & 0.0610 & 0.1680 & 0.2220 & 0.007 & 0.0230 & 0.0475 & 0.0530 & 0.003 \\
\hline

\% \text{Change} & \textbf{-8.5} & \textbf{-8.4} & \textbf{-9.2} & - & \textbf{-8.1} & \textbf{-9.1} & \textbf{-9.7} & - & \textbf{-5.3} & \textbf{-5.4} & \textbf{-6.1} & - & \textbf{-5.9} & \textbf{-6.2} & \textbf{-6.7} & - \\
\hline
\end{tabular}
\end{table*}

\subsection{Overall Performence(RQ1)}\label{sec:Overall}
To answer \textbf{RQ1}, we conducted a comprehensive performance comparison against a wide range of baselines. To ensure statistical robustness, all experiments for all methods were run 3 times using different random seeds (1024, 2046, 4096). As shown in Table~\ref{tab:performance}, we report the average performance of these runs and include the standard deviation (Std) for the R@20 metric to demonstrate result stability. Following the reviewer's valuable suggestions, we have enriched our comparison by adding four new methods: (1) two sampling-based variants, RanSAM and MajSAM, which serve as ablation studies on 80\% data, and (2) two recent compress-based baselines, UnKD (2023) and GASD (2024). 

The results are shown in Table~\ref{tab:performance} and the analysis is listed below:

\begin{itemize}[leftmargin=*] 
\item [(i)] It is critical to note that DemoRec's primary objective is to optimize the trade-off between efficiency and accuracy, not solely to achieve the absolute highest recommendation performance. In our primary comparison, the SOTA baseline (\eg LightGCL) is trained on the \textbf{full 100\%} of the original graph data, whereas DemoRec operates on a condensed graph representing only 80\% of the data size. The fact that DemoRec, despite using 20\% less data, achieves highly competitive performance (a minimal $\approx$ 5\% to 10\% drop, as shown in the \% Change row) is the key finding. This demonstrates that substantial space and time savings (as shown in Section IV.E) can be achieved with only a minor and acceptable loss in accuracy.
\item [(ii)] DemoRec's superiority over sampling-based variants is clear. As shown in Table~\ref{tab:performance}, DemoRec significantly outperforms both MajSAM (Majority Sampling) and RanSAM (Random Sampling). This confirms that naively discarding 20\% of nodes (by degree or randomly) results in critical information loss. In contrast, our condensation framework effectively synthesizes information from the \textit{entire} original graph into a compact, superior-performing subgraph.
\item [(iii)] \textbf{Comparison with other compression-based methods.} We compare DemoRec against two recent compression-based baselines, UnKD and GASD, which are evaluated at the same 80\% data ratio (a 20\% reduction). As shown in the table, all methods operating on 80\% of the data (including ours) expectedly underperform the state-of-the-art model (LightGCL) trained on the full (100\%) dataset. This performance gap highlights a key limitation of traditional knowledge distillation (KD) methods like UnKD and GASD: while attempting to compress, they often focus on matching node-level outputs, consequently damaging or losing critical graph topological information in the process. Furthermore, many of these compression techniques are extremely time-consuming, as their distillation or complex processing (\eg, in GASD) still requires extensive computation over the full, large-scale original graph. In contrast, our condensation framework is specifically designed to preserve this vital topological structure via its alignment modules, resulting in a more efficiently-generated condensed graph that better retains structural integrity.
\item [(iv)] Our method on average outperforms traditional CF-based models (\eg NGCF, GF-CF), achieving superior performance while also saving space. This demonstrates that our approach not only reduces the space complexity but also enhances the recommendation accuracy, surpassing some of the existing recommendation algorithms.
\end{itemize}

%\yhr{Why is the analysis about other baselines deleted? Only two points listed here are obviously insufficient.}
These advantages underscore the potential of our method to improve both efficiency and effectiveness in recommender systems.
% These findings collectively demonstrate that DemoRec not only effectively condenses large-scale graphs but also ensures competitive performance across different model types \yhr{what model types?}, making it a promising framework for scalable and efficient recommender systems.

\subsection{Ablation Study(RQ2)}\label{sec:ablation}
%\yhr{This section is important and should be placed before Section 3.3.}
%\yhr{Is this RQ5?}

To answer \textbf{RQ2}, we conducted an ablation study in Gowalla by removing this component from our DemoRec framework. The purpose of this study was to investigate how the BSL impacts the condensed graph and its resulting recommendation performance. Specifically, the BSL ensures that the condensed graph retains its user-item bipartite structure, avoiding the creation of redundant user-user and item-item connections. Such redundant connections introduce noise into the recommendation process, leading to degraded performance. 

The results of this ablation study are shown in Fig.~\ref{fig:bipartite_loss}, where the performance of DemoRec and Non-Bi DemoRec is compared across Precision, Recall, and NDCG. From the figure, it is evident that DemoRec significantly outperforms Non-Bi DemoRec on all metrics. This Observation can be attributed to the following factors:

\begin{figure}[t]
\centering
\includegraphics[width=1\columnwidth]{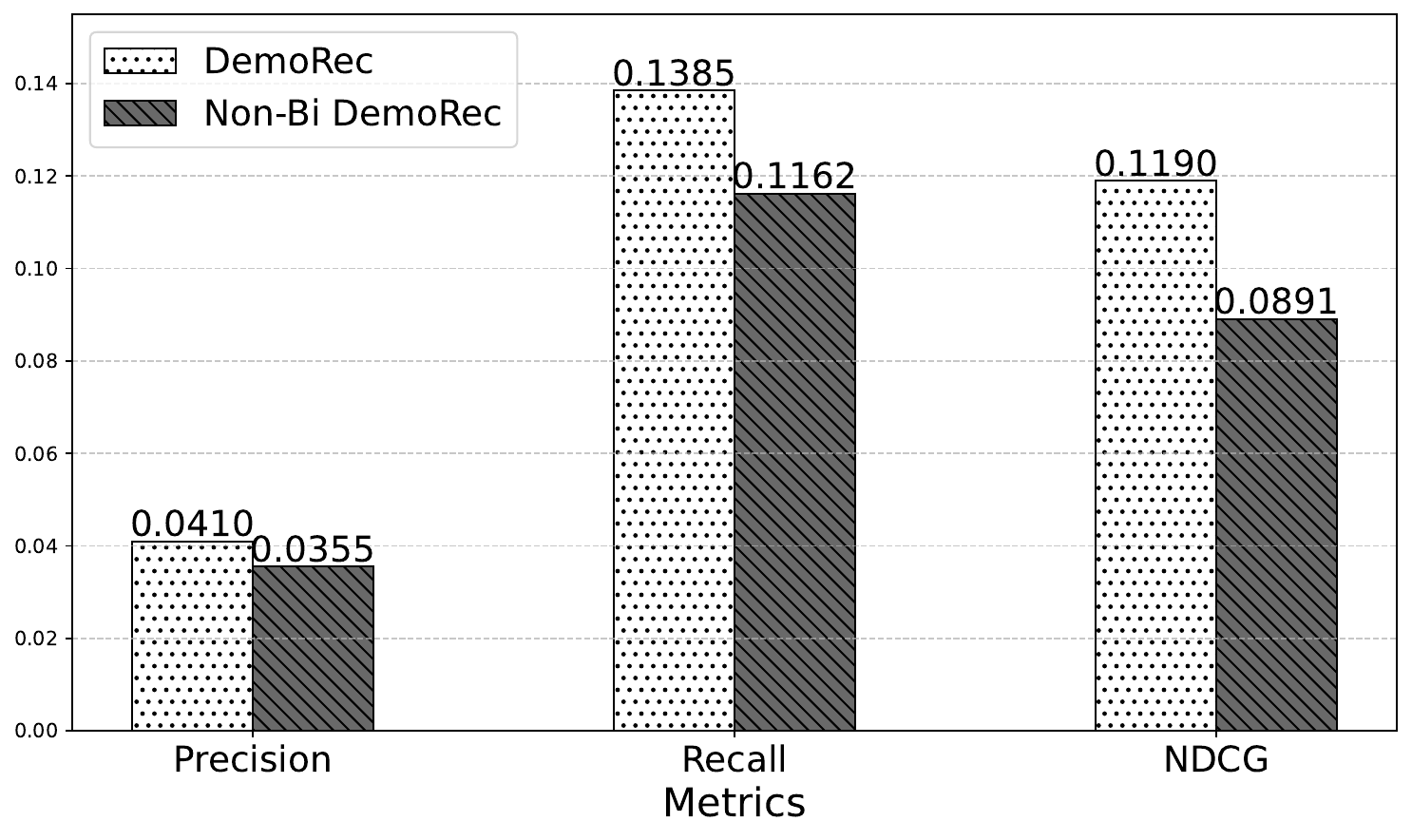}
\caption{Performance comparison between DemoRec and Non-Bi DemoRec on Precision, Recall, and NDCG.}
\label{fig:bipartite_loss}
\end{figure}

\begin{itemize}[leftmargin=*] 

\item [(i)] The removal of the BSL allows the condensation process to create non-bipartite connections (\ie user-user and item-item links exist), which are redundant %\yhr{do not use meanningful} 
in the context of user-item interaction graphs. These redundant connections introduce irrelevant information into the recommendation process, directly leading to performance degradation.

\item [(ii)] The BSL ensures that the semantic integrity of the graph is preserved during condensation. Without this constraint, the condensed graph no longer accurately reflects the original user-item relationships, causing the model to struggle in capturing meaningful patterns for recommendation.

\item [(iii)] As shown in Fig.~\ref{fig:bipartite_loss}, the performance gap between models with and without the BSL increases as the graph condensation ratio becomes higher. This trend suggests that preserving the bipartite structure is particularly critical in scenarios where graph information is already reduced, as even small amounts of redundant connections can disproportionately affect performance under high compression.

These findings highlight the necessity of the BSL in ensuring that the condensed graph remains a faithful representation of the original user-item interaction graph. By maintaining the user-item bipartite structure, the  prevents the introduction of noisy connections and enhances the overall robustness of the DemoRec framework.
\end{itemize}

\subsection{Component Analysis(RQ3)}\label{sec:component}

To address \textbf{RQ3}, we conducted a series of experiments to analyze the impact of different GNN architectures (GCN, GAT, and GraphSAGE) when applied within our DemoRec framework. These architectures were chosen to evaluate the adaptability and robustness of our graph condensation approach across backbone models with varying levels of complexity. The condensation ratio refers to the proportion of the original graph that remains after condensation. 
Furthermore, we analyze the impact of different matching strategies for $\mathcal{L}_{match}$, which is a critical component of our condensation framework. As $\mathcal{L}_{relay}$, $\mathcal{L}_{match}$, and $\mathcal{L}_{rec}$ are all essential, co-dependent components, a simple removal-based ablation is not feasible. Instead, we conduct a replacement analysis comparing the three primary matching methods discussed in our framework: Gradient (D.(Grad.)), Trajectory (D.(Traj.)), and Distribution matching (D.(Dist)).

The results are shown in Fig.~\ref{fig:Component_matching} and Fig.~\ref{figure:RPD}, the related analysis is summarized as follows:

\begin{itemize}[leftmargin=*] 

\item [(i)] We observe that, for the same graph condensation ratio, the performance degradation of GAT and GraphSAGE is significantly lower than that of GCN when compared to their respective performance at $\alpha = 1$. This indicates that our graph condensation method is particularly effective for more complex GNN models, as they can better retain information from the condensed graph. Specifically, the advanced architectural mechanisms of GAT and GraphSAGE enable them to adapt more effectively to the reduced graph structure, whereas GCN, which relies primarily on simple aggregation mechanisms, struggles to maintain performance under condensation.

\item [(ii)] Furthermore, as shown in Fig.~\ref{figure:RPD}, the overall performance degradation slope of GAT is lower than that of GraphSAGE. This can be attributed to GAT’s attention mechanism, which selectively emphasizes important edges or neighbors during message passing. This selective emphasis enhances scalability and robustness during training, and the condensation process further reinforces this capability, leading to a less pronounced performance degradation in GAT compared to GraphSAGE.

% \begin{table}[t]
% \centering
% \caption{Relative Performance Degradation (RPD) Across GNN Architectures. \yhr{It is unnecessary to list this table. It can be integrated into Fig. 2 with RPD marked on each data point.}}
% \label{table:RPD}
% \renewcommand{\arraystretch}{1.5}
% \resizebox{\columnwidth}{!}{
% \begin{tabular}{l|c|c|c}
% \hline
% \textbf{Model} & \textbf{condensation ratio (\%)} & \textbf{Performance Loss (\%)} & \textbf{RPD} \\
% \hline
% DemoRec (GCN)       & 50                     & 20                    & 0.4  \\
% DemoRec (GraphSAGE) & 50                     & 10                    & 0.2  \\
% DemoRec (GAT)       & 50                     & 12                    & 0.24 \\
% \hline
% \end{tabular}
% }
% \end{table}

\item [(iii)] Across all the backbone models, we found an insightful trend: as the graph condensation ratio decreased, the slope became lower.
This is evident from Fig.~\ref{figure:RPD}, where the slope of the performance loss curve becomes flatter at a lower condensation ratio. 
These findings highlight the versatility and robustness of DemoRec when applied to different GNN architectures. Our method not only performs better with more complex models like GAT and GraphSAGE but also shows increasing efficiency at a lower condensation ratio, making it a compelling solution for scalable and efficient graph-based recommender systems. This suggests that the condensation process not only compresses the graph but also removes redundant information, enabling the model to focus on the most relevant structural features.

\item [(iv)] The results (shown in Fig.~\ref{fig:Component_matching}) indicate that Trajectory matching yields the best recommendation performance. This is attributed to its ability to capture the complete training trajectory, making it the most effective approach for preserving critical information. Gradient matching follows, lagging slightly in performance, as it primarily aligns the optimization direction with that of the original graph. Distribution matching performs the worst. This is because it is the simplest method, essentially capturing only a static snapshot of the model's state. It only aligns the embedding distribution statistics after the GNN forward pass, completely ignoring the crucial dynamics involved in backpropagation and parameter optimization.

\begin{figure}[t]
\centering
\includegraphics[width=1\columnwidth]{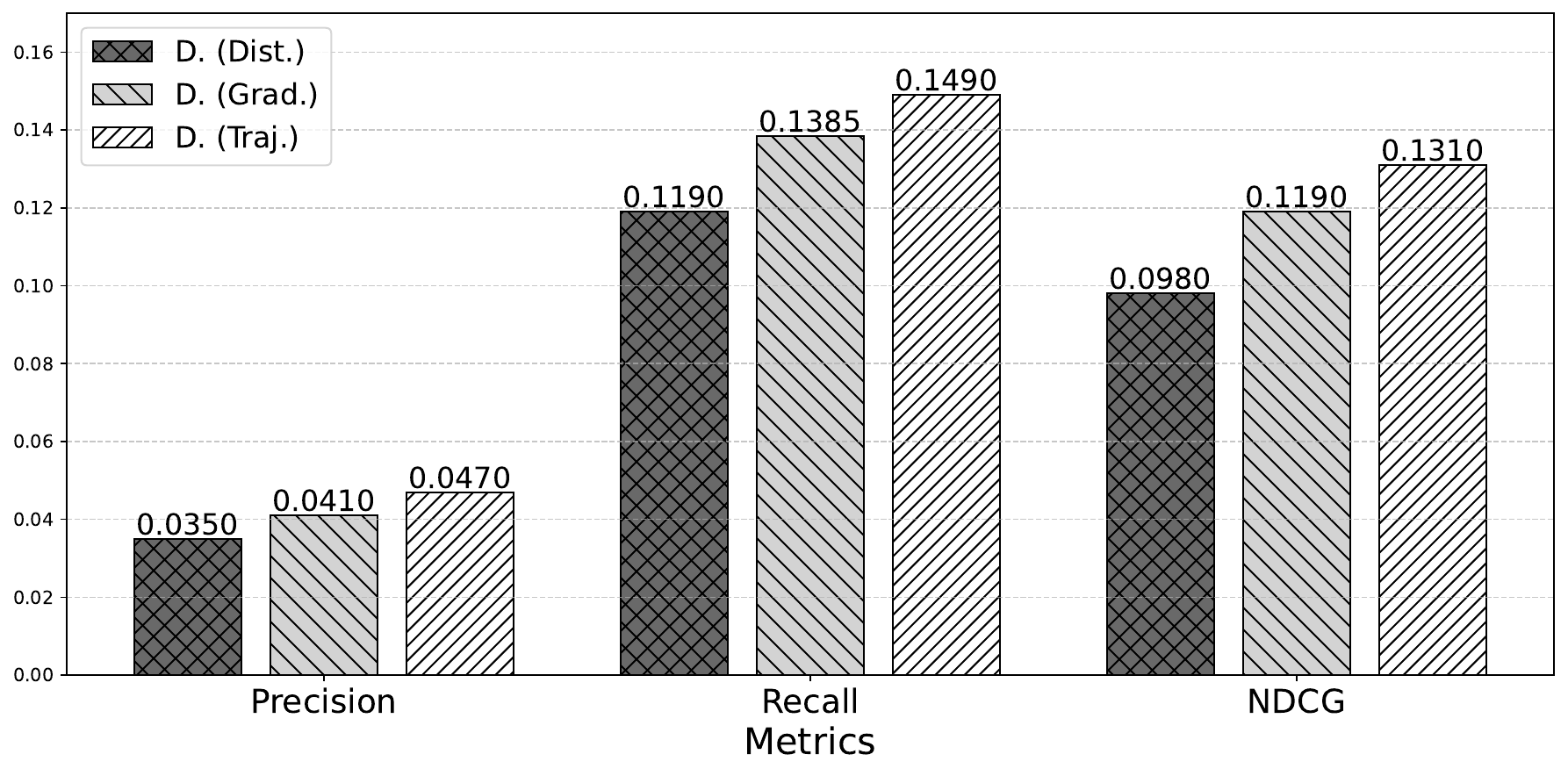}
\caption{Component Analysis on different matching components (D. (Grad.), D. (Traj.), and D. (Dist.))}
\label{fig:Component_matching}
\end{figure}

\end{itemize}
\begin{figure}[t]
\centering
\includegraphics[width=0.9\columnwidth]{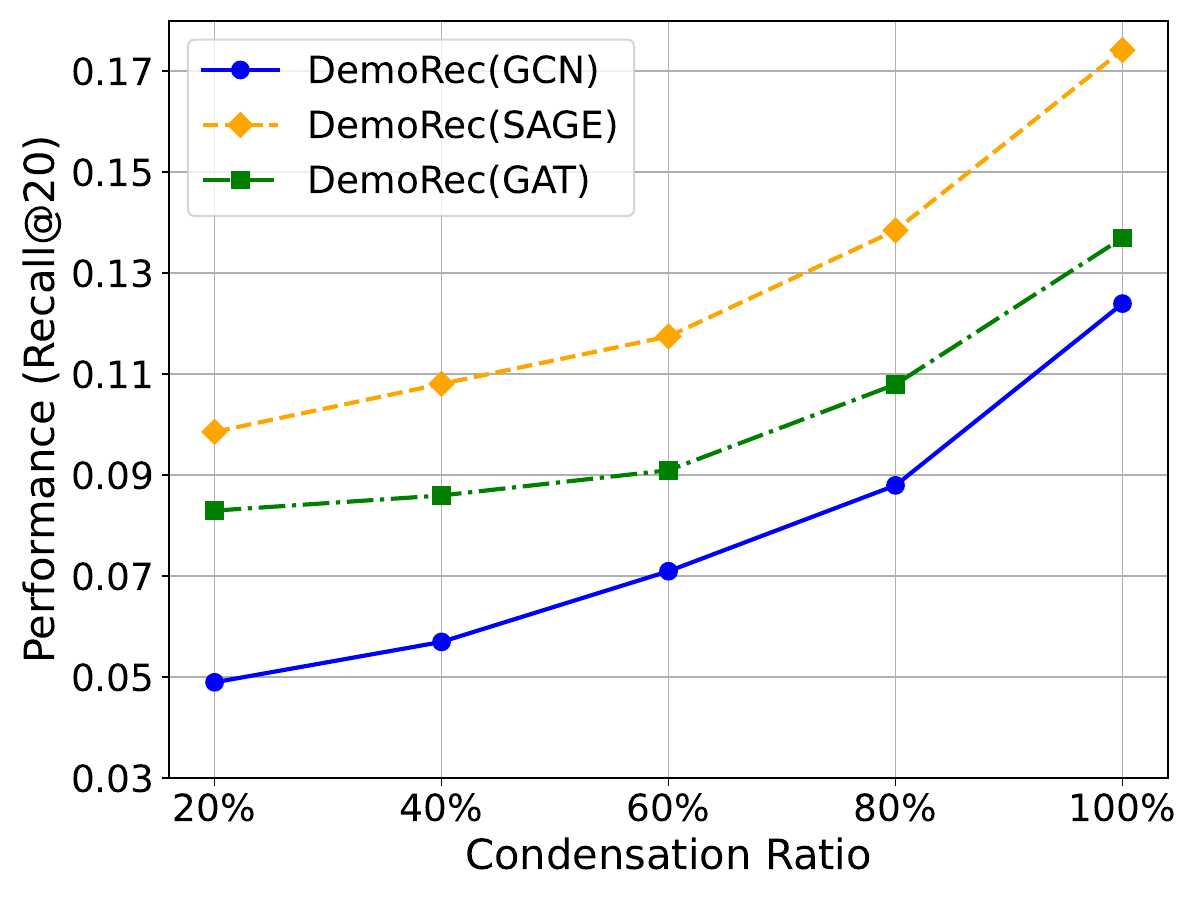}
\caption{Component Analysis with three backbone GNNs.}
\label{figure:RPD}
\end{figure}

These results above highlight the superiority of DemoRec in maintaining strong performance across different GNN architectures, with particularly promising results for more complex models and higher condensation ratios. The flexibility and adaptability of DemoRec ensure its suitability for large-scale graph scenarios.

% --- 使用 table* 环境使其跨越两栏 ---
\begin{table*}[t]
\centering
\caption{Efficiency Study Across Backbone Models. Time (training time per epoch) and Space (GPU memory consumption) are measured on the Gowalla dataset.}

\label{table:efficiency_study} 
\renewcommand{\arraystretch}{1.5}

\setlength{\tabcolsep}{2.1pt} 

\begin{tabular}{l|rrr|rr|rrr|rrrr}
\toprule
& \multicolumn{3}{c|}{\textit{GNN Baselines (100\%)}} 
& \multicolumn{2}{c|}{\textit{Compress-based (80\%)}}
& \multicolumn{3}{c|}{\textit{Our Method (Backbones)}}
& \multicolumn{4}{c}{\textit{Our Method (Ablation)}} \\
\cline{1-1}

\diagbox{\textbf{Metric}}{\textbf{Method}} & LightGCN & UltraGCN & LightGCL 
              & UnKD & GASD
              & D. (GCN) & D. (SAGE) & D. (GAT)
              & D. (Grad.) & D. (Traj.) & D. (Dist.) & D. w/o BSL \\
\midrule

\textbf{Time (s)} & 7.01 & 12.45 & 21.30
                & 35.82 & 52.41
                & 7.32 & 6.50 & 9.84
                & 7.32 & 9.15 & 7.03 & 7.17 \\
                
\textbf{Space (MB)} & 142.5 & 160.2 & 190.8
                 & 220.4 & 245.1
                 & 138.3 & 165.1 & 172.3
                 & 138.3 & 145.0 & 137.5 & 138.0 \\
\bottomrule
\end{tabular}
\end{table*}

\subsection{Efficiency Study(RQ4)}\label{sec:efficiency}
To answer \textbf{RQ4}, we conducted a comprehensive comparative study between GNN and compress-based baseline methods and our DemoRec framework (instantiated with GCN, SAGE, and GAT backbones). The purpose of this experiment was to assess the differences in time and space requirements, focusing on the computational and storage efficiency of our method. Specifically, time efficiency was measured by the training time per epoch in seconds. Space efficiency was measured by the GPU memory consumption during training in megabytes (MB). For fair comparisons, all experiments were conducted on the Gowalla dataset, and the hidden size of each model was fixed and set to 64. We exclude traditional CF-based models (e.g., GF-CF) from this comparison. Their significantly lower performance (shown in Table \ref{tab:performance}) and different computational paradigms (i.e., not relying on GNN-style message passing and backpropagation) make a direct efficiency comparison inequitable. We also clarify that the core graph "mapping" (i.e., the generation of the condensed adjacency matrix $\mathbf{A}'$) is part of the graph condensation process, which is a pre-computation step before the per-epoch training measured here. Therefore, it does not introduce latency to the training times reported in Table~\ref{table:efficiency_study}.
The results of the efficiency study are shown in Table~\ref{table:efficiency_study} and summarized as follows:

The results of the efficiency study are shown in Table~\ref{table:efficiency_study} and summarized as follows:

\begin{itemize}[leftmargin=*] 
\item [(i)] \textbf{Comparison with GNN Baselines (100\% Data): This is the key comparison. As shown in Table \ref{table:efficiency_study}, all variants of our DemoRec framework (operating on 80\% data) are significantly more efficient than the 100\%-data GNN baselines. DemoRec (all variants) overwhelmingly outperforms advanced GNNs like LightGCL, which incurs heavy overhead from its contrastive learning mechanism. Most importantly, our most time-efficient variant, DemoRec (SAGE), is faster (6.50s) than the highly-optimized and efficient baseline LightGCN (7.01s). This demonstrates a clear efficiency win against the state-of-the-art.}

\item [(ii)] \textbf{Comparison with Compress-based Baselines (80\% Data): Our method also demonstrates superior efficiency against other compress-based techniques operating on the same 80\% data. The baseline methods, UnKD and GASD, are extremely time-consuming. This is because their distillation processes still require extensive computation and message passing over the \textit{full, large-scale original graph}. GASD is particularly slow, as its methodology involves complex and costly operations in hyperbolic space. In contrast, our DemoRec, once condensed, trains \textit{only} on the small graph, leading to a massive speedup in the training phase.}

\item [(iii)] \textbf{DemoRec Backbone Analysis:} Our framework inherits and enhances the characteristics of its backbones. DemoRec (SAGE) achieves the best time efficiency, building on SAGE's fast neighborhood sampling. DemoRec (GCN) achieves the lowest overall space consumption (138.3 MB), making it the most memory-efficient option, outperforming even the lightweight LightGCN baseline. DemoRec (GAT) is naturally slower and more memory-intensive due to its attention mechanism.

\item [(iv)] \textbf{Matching Method Latency: We analyzed the efficiency trade-offs of the three matching strategies, which present a clear trade-off between performance and efficiency. Distribution matching (D.(Dist.)) is the fastest (7.03s), as it only requires a GNN forward pass, but (as shown in our component study) yields the lowest performance. Trajectory matching (D.(Traj)) is the slowest (9.15s), as it must unroll a full K-step training process to compute meta-gradients, but it achieves the best performance by capturing the full learning dynamic. Our chosen method, Gradient matching (D.(Grad.)) (7.32s), offers a balanced trade-off, being much faster than Trajectory matching as it only requires a single gradient computation.}

\item [(v)] \textbf{BSL Latency:} We conducted an additional efficiency analysis by comparing DemoRec with and without the Bipartite Structure Loss (BSL) calculation on the Gowalla dataset. The results show that BSL adds only a marginal increase in training time per epoch (approx. 2.1\%, comparing DemoRec (GCN) at 7.32s to 7.17s w/o BSL), which is minimal compared to the efficiency gains from the graph condensation itself. 
\end{itemize}

These findings highlight that DemoRec not only provides a strong accuracy-efficiency trade-off but also achieves superior or competitive time and space efficiency compared to current state-of-the-art baselines like LightGCN and LightGCL.

\subsection{Hyper-Parameter Study(RQ5)}\label{sec:parameter}
%\yhr{Is this RQ4?}

\begin{figure}[t]
\centering
\includegraphics[width=1\columnwidth]{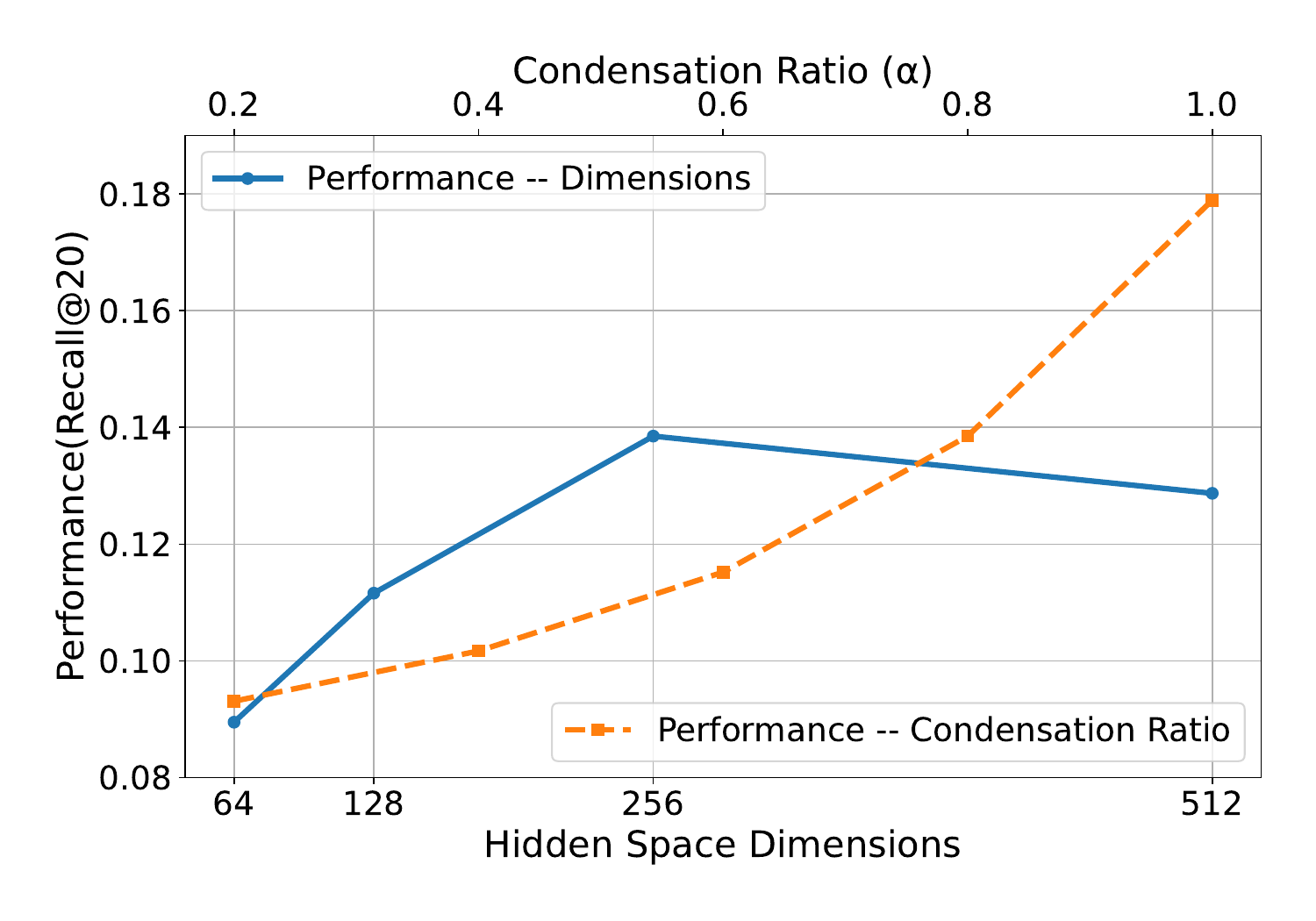}
\caption{Hyper-parameter study of hidden space dimensions and condensation ratio $\alpha$.}
\label{figure:hyper_parameter}
\end{figure}

\begin{figure}[t]
\centering
\includegraphics[width=1\columnwidth]{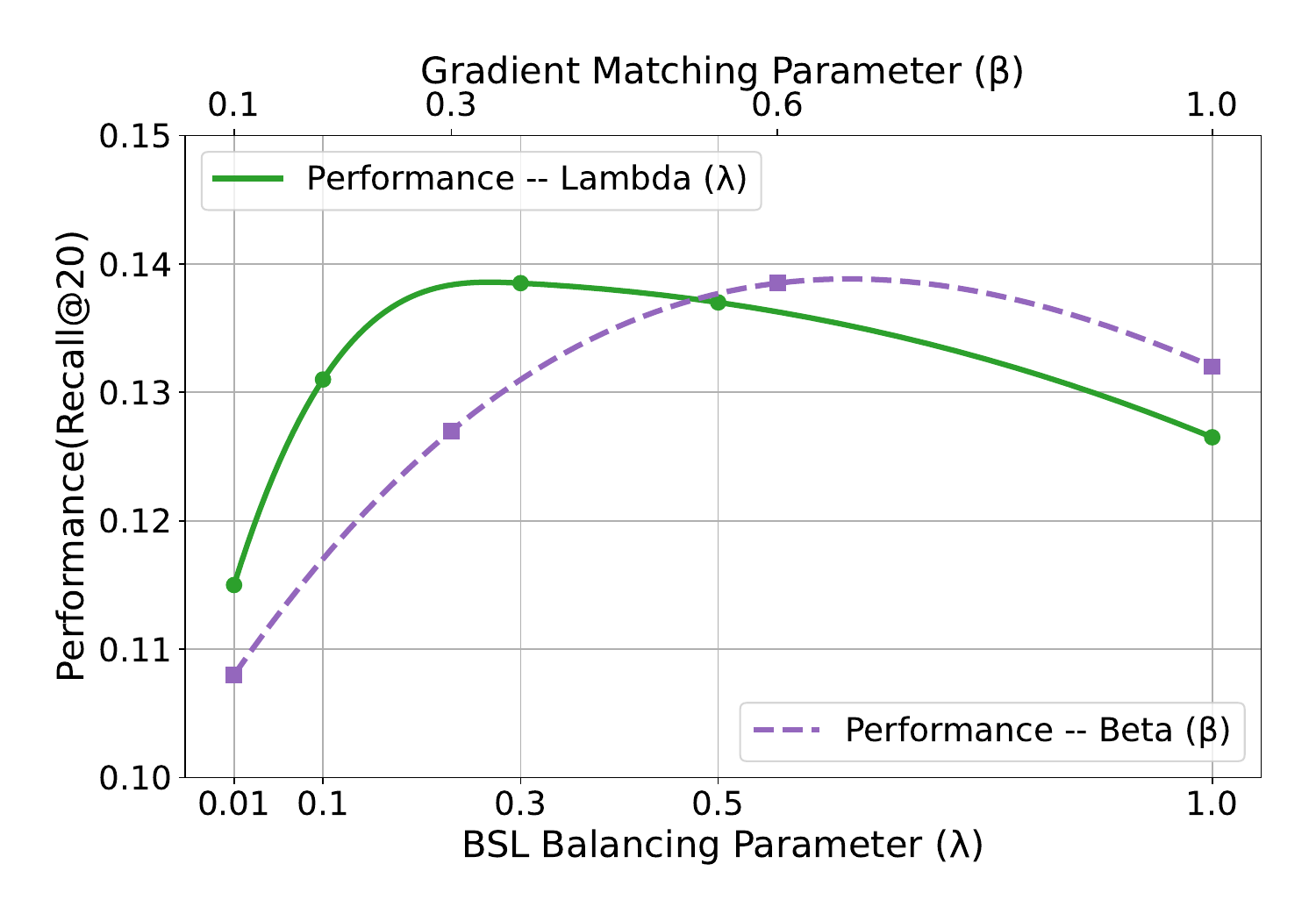}
\caption{Hyper-parameter study of BSL balancing parameter $\lambda$ and gradient matching parameter $\beta$.}
\label{figure:hyper_lambda_beta}
\end{figure}

To answer \textbf{RQ5}, we conducted four hyper-parameter experiments to evaluate the performance of our DemoRec framework under various configurations. We focused on four key hyper-parameters: the hidden space dimensions, the condensation ratio $\alpha$, the BSL balancing parameter $\lambda$, and the gradient matching parameter $\beta$. The selection of the optimal hidden space dimensions is imperative for ensuring the efficacy and efficiency of the model. In this study, four values (64, 128, 256, and 512) were evaluated to ascertain the most suitable dimension for the hidden space.
Additionally, we varied the condensation ratio $\alpha$ to assess its impact on model performance, using a range of values to explore how different levels of graph condensation affect performance.
The results of these experiments are summarized in Fig. \ref{figure:hyper_parameter} and \ref{figure:hyper_lambda_beta}, which plots the recommendation performance against each of these key hyper-parameters.

\begin{itemize}[leftmargin=*]
\item [(i)] In the first experiment, we evaluated the performance of DemoRec with different hidden space dimensions. We found that DemoRec performed best with a hidden space dimension of 256. As the hidden space size increased, the model showed improved performance, up to a point where the additional capacity did not lead to significant gains. This suggests that a hidden space size of 256 strikes the satisfying balance between model complexity and performance, enabling DemoRec to capture the relevant features effectively without overfitting or underfitting.

\item [(ii)] The second experiment focused on varying the condensation ratio $\alpha$. 
As the condensation ratio increased, we observed that the curve slope decreased, 
indicating that the model retained more useful information as $\alpha$ increased. 
Specifically, the performance decline became slower with a higher condensation ratio, suggesting that DemoRec can effectively compress large graphs while still preserving enough information for accurate recommendations. This phenomenon was most noticeable in scenarios with limited data, where our model showed a clear advantage. The increasing preservation of information at a higher condensation ratio allowed DemoRec to perform well even when the graph was highly compressed, making it particularly suited for environments with limited data. 

\item [(iii)] The third experiment investigates the sensitivity of the BSL balancing parameter $\lambda$. This parameter manages the trade-off between preserving the bipartite structure and overall recommendation accuracy. We varied $\lambda$ (\eg in the range of [0.01, 0.1, 0.3, 0.5, 1.0]) and observed its impact on performance. The results indicate that as $\lambda$ increases, the model more strictly enforces the user-item bipartite structure, which is crucial for structural integrity. However, an overly high value (\eg $\lambda > 0.3$) can lead to a decline in recommendation performance, as the model over-prioritizes structural constraints at the expense of capturing nuanced user-item preferences. Our experiments consistently show that $\lambda = 0.3$ achieves the optimal balance, yielding the best recommendation results.

\item [(iv)] The fourth experiment examines the gradient matching parameter $\beta$ (Eq. 22). This parameter balances the importance of the gradient matching loss during the condensation stage. We tested values in the range of [\eg 0.1, 0.3, 0.6, 1.0]. A small $\beta$ results in poor alignment, as the condensation process fails to capture the learning dynamics of the original graph. Conversely, a very large $\beta$ over-emphasizes gradient alignment at the expense of other objectives. Our study finds that $\beta = 0.6$ provides the best trade-off, leading to the highest recommendation accuracy.

\end{itemize}

These results demonstrate that DemoRec's performance is sensitive to its key hyper-parameters. Optimal performance was achieved with a hidden space size of 256, a sufficiently high condensation ratio $\alpha$, a BSL parameter $\lambda=0.3$, and a gradient matching parameter $\beta=0.6$. This highlights the robustness of our framework, as it can maintain strong performance even when faced with large graph compression and small datasets.

\section{Related Work}\label{sec:related_work}
%\yhr{Use one sentence to simply describe the following content.}
In this section, we provide an overview of two key research areas that closely relate to our study: Graph Condensation and Graph Recommendation. These topics serve as the foundation for understanding the context and contributions of our work, offering insights into how our proposed methods build upon and extend existing efforts in the field.

\subsection{Graph Condensation}

Graph condensation is an emerging subfield of data distillation that focuses on compressing graph structures while preserving their essential properties for downstream tasks. Inspired by dataset distillation in computer vision~\cite{radosavovic2018data,chen2022dearkd}, graph condensation methods aim to generate a smaller, synthesized graph that mimics the original graph's behavior. Unlike traditional graph reduction techniques such as sparsification~\cite{bravo2019unifying} and coarsening~\cite{cai2021graph}, which rely on heuristics to prune edges or merge nodes, graph condensation employs optimization-based approaches to directly learn a smaller graph that approximates the original graph's predictive performance.

Recent studies in graph condensation have demonstrated its potential to reduce graph size while maintaining high performance. For instance, GCond~\cite{jin2021graph} proposed a framework for condensing graphs through gradient matching, achieving significant reductions in graph size without degrading GNN performance. The field has since expanded rapidly, exploring diverse techniques such as contrastive learning~\cite{gao2025contrastive} and policy gradient estimation~\cite{wu2024condensing}, including applications tailored for recommendation~\cite{wu2025leveraging}. Other recent works have focused on critical aspects like robustness~\cite{gao2025robgc} and security~\cite{wu2025backdoor}. Similarly, methods like GCond~\cite{liu2023graph} introduced more sophisticated optimization objectives to improve the fidelity of the condensed graph, enabling its application to larger and more complex datasets.

Despite these promising developments, current graph condensation techniques still face notable challenges, particularly when applied to specialized graph structures like bipartite user-item interaction graphs. Bipartite graphs, characterized by their division into two distinct node sets (\eg users and items) with edges existing only between the sets and not within them, possess unique structural properties that many existing methods struggle to preserve. Studies such as~\cite{he2016birank} and~\cite{li2021bipartite} highlight how traditional condensation approaches often overlook these characteristics, such as the absence of intra-group connections (\eg user-to-user or item-to-item links). As a result, the condensed graphs produced by these methods may fail to accurately reflect the original bipartite structure, leading to degraded performance in tasks like recommendation or ranking.

Our work directly tackles this limitation by introducing a novel contribution, the BSL. This specialized loss function is designed to explicitly enforce the preservation of the bipartite nature of the graph during the condensation process.

\subsection{Graph Recommendation}

Graph-based methods have achieved significant progress in recommender systems in recent years~\cite{wu2022graph,10113669}. Traditional approaches, such as collaborative filtering (CF)~\cite{he2017neural,wang2019neural,shen2021powerful,wang2020disentangled,doi:10.1504/IJAACS.2023.131069,doi:10.1504/IJAACS.2023.131079} and content-based recommendations\cite{reddy2019content,javed2021review,doi:10.1504/IJAACS.2022.127411}, primarily rely on static user and item attributes, as well as explicit user feedback. However, these methods struggle to model complex interactions between entities, or fail to fully utilize the topological properties of the graph itself.

To address this limitation, researchers have developed recommendation systems based on Graph Neural Networks (GNNs)~\cite{wu2021self, chen2023heterogeneous, yu2022graph, cai2023lightgcl, peng2023dual}. GNNs efficiently process graphs of varying sizes by iteratively propagating and aggregating semantic information from topological neighbors, thereby learning low-dimensional embeddings~\cite{wu2022graph}. For instance, LightGCN~\cite{he2020lightgcn} enhances recommendation performance by simplifying the graph neural network (GNN) architecture. Specifically, it removes nonlinear activation functions and redundant weight matrices, leading to more efficient and effective learning. Additionally, MultiCSR~\cite{chen2024multi} introduces a multi-view contrastive learning framework that integrates users' social networks and item knowledge graphs, enhancing user preference modeling through contrastive learning. 
Despite these advancements, GNN-based recommendation models are not without their challenges, particularly when applied to large-scale datasets characterized by extensive user-item interactions. The message-passing paradigm central to GNNs, while effective in capturing neighborhood information, often introduces substantial computational overhead, especially as graph size increases~\cite{hu2020heterogeneous,gao2018large}. This scalability issue can hinder real-world deployment, where efficiency is paramount. To address this, recent work such as HGNR~\cite{liu2020heterogeneous} introduces a higher-order bipartite tower structure within a heterogeneous graph convolutional network (GCN). By explicitly modeling higher-order interactions in heterogeneous graphs, HGNR mitigates inefficiencies inherent in traditional GNNs, thereby improving the quality of user and item representations while offering a more scalable solution for recommendation systems.

Nonetheless, these methods remain computationally demanding and may exhibit suboptimal performance in datasets with a vast number of entities. To overcome these challenges, our approach leverages graph condensation to improve efficiency while maintaining recommendation effectiveness. By condensing the original graph into a smaller, yet representative structure, our approach substantially reduces the computational burden while preserving the essential properties needed for effective recommendations.

\section{Conclusion}
This paper introduces DemoRec, an innovative framework that utilises graph compression techniques to improve the scalability and accuracy of large-scale user-item interaction graph recommendation systems. By generating representative nodes for users and items, DemoRec constructs a compact interaction graph, effectively reducing the graph size and computational complexity while avoiding the problem of over-reliance on higher-order information. Additionally, we propose the Bipartite Structure Loss (BSL), which enforces structural consistency in the condensed graph by preserving critical user-item relationships, with theoretical proofs and ablation studies confirming its effectiveness in enhancing recommendation performance. Experiments on multiple public graph datasets show that DemoRec is not only comparable to existing SOTA methods in terms of recommendation performance, but also significantly outperforms all existing SOTA baselines in terms of computational efficiency. However, DemoRec's reliance on static graphs and the computational overhead of the condensation stage itself present limitations. Future work will therefore aim to develop an incremental version for dynamic graphs and explore more computationally efficient condensation algorithms.

% \subsection*{ACKNOWLEDGEMENTS}
% This research was partially supported by APRC - CityU New Research Initiatives (No.9610565, Start-up Grant for New Faculty of City University of Hong Kong), SIRG - CityU Strategic Interdisciplinary Research Grant (No.7020046, No.7020074), HKIDS Early Career Research Grant (No.9360163), Huawei (Huawei Innovation Research Program) and Ant Group (CCF-Ant Research Fund), and Key Laboratory of Smart Education of Guangdong Higher Education Institutes, Jinan University (2022LSYS003).
\bibliographystyle{ieeetr}

\bibliography{Reference}
%\clearpage
%\appendix
%\input{6Appendix}

\begin{IEEEbiography} [{\includegraphics[width=1in,height=1.25in,clip,keepaspectratio]{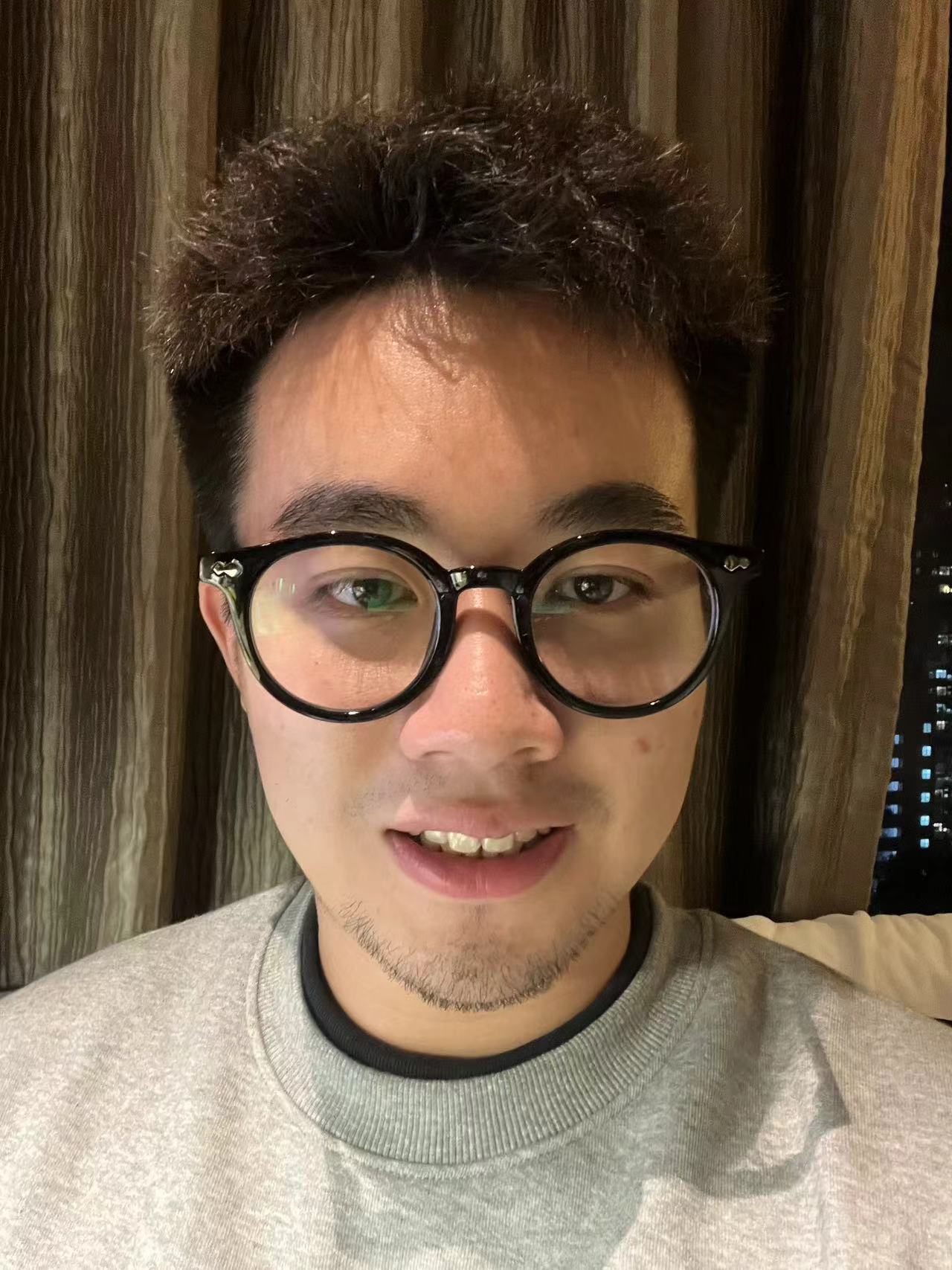}}] {Jiahao Liang} received the M.S. degree from City university of Hong Kong in 2023. He is currently pursuing the Ph.D. degree at the School of Future Technology, South China University of Technology, China. His research interests include machine learning and graph data mining.
\end{IEEEbiography}

\begin{IEEEbiography}[{\includegraphics[width=1in,height=1.25in,clip,keepaspectratio]{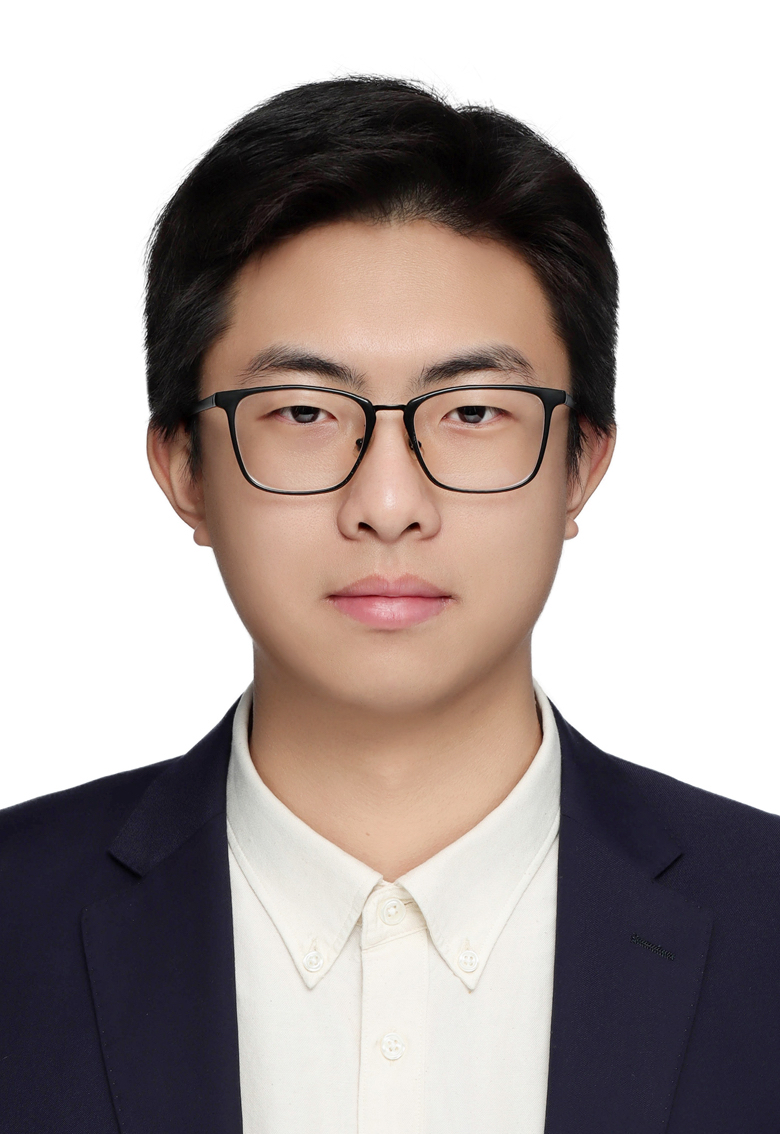}}]{Haoran Yang} 
obtained his Ph.D. degree at the School of Computer Science (SoCS) of the University of Technology Sydney (UTS) in 2025, supervised by Prof. Guandong Xu. He received his B.Sc. in Computer Science and Technology and B.Eng.; Minor in Financial Engineering from Nanjing University (NJU) in 2020. His research interests include, but are not limited to, graph data mining and its applications.
\end{IEEEbiography}

\begin{IEEEbiography}[{\includegraphics[width=1in,height=1.25in,clip,keepaspectratio]{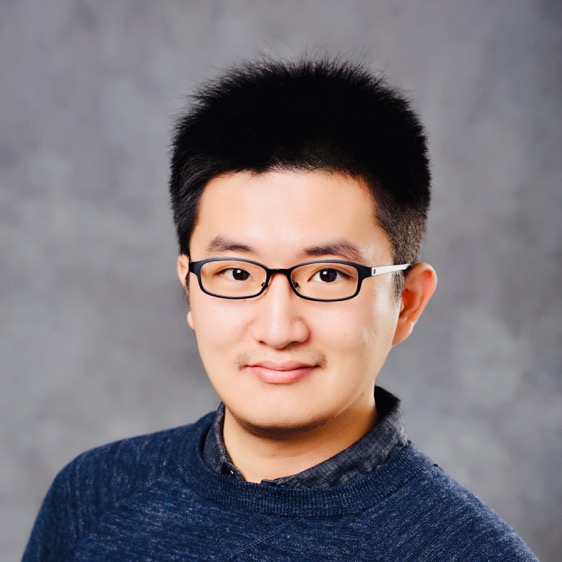}}]{Xiangyu Zhao} is a tenured associate professor (effective from July 2025) of Data Science at City University of Hong Kong (CityU). He worked as an assistant professor at CityU from Sep 2021, and then got an early promotion with tenure. Prior to CityU, he completed his Ph.D. under Prof. Jiliang Tang at MSU, his M.S. under Prof. Enhong Chen at USTC, and his B.Eng. under Prof. Tao Zhou and Prof. Ming Tang at UESTC. He has published more than 100 papers in top conferences and journals. His research has been awarded ICDM'22 and ICDM'21 Best-ranked Papers and Global Top 25 Chinese New Stars in AI (Data Mining). He is a member of the founding academic committee of MLNLP, the largest Chinese AI community with millions of subscribers.
\end{IEEEbiography}

\begin{IEEEbiography} [{\includegraphics[width=1in,height=1.25in,clip,keepaspectratio]{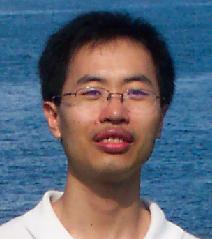}}] {Zhiwen Yu} is a Professor in School of Computer Science and Engineering, South China University of Technology, China.  Dr. Yu serves as an associate editor of IEEE Transactions on Systems, Man, and Cybernetics: Systems. He received the PhD degree from City University of Hong Kong in 2008. Dr. Yu has published more than 150 referred journal papers and international conference papers, including 50 IEEE Transactions papers.
\end{IEEEbiography}

\begin{IEEEbiography} [{\includegraphics[width=1in,height=1.25in,clip,keepaspectratio]
{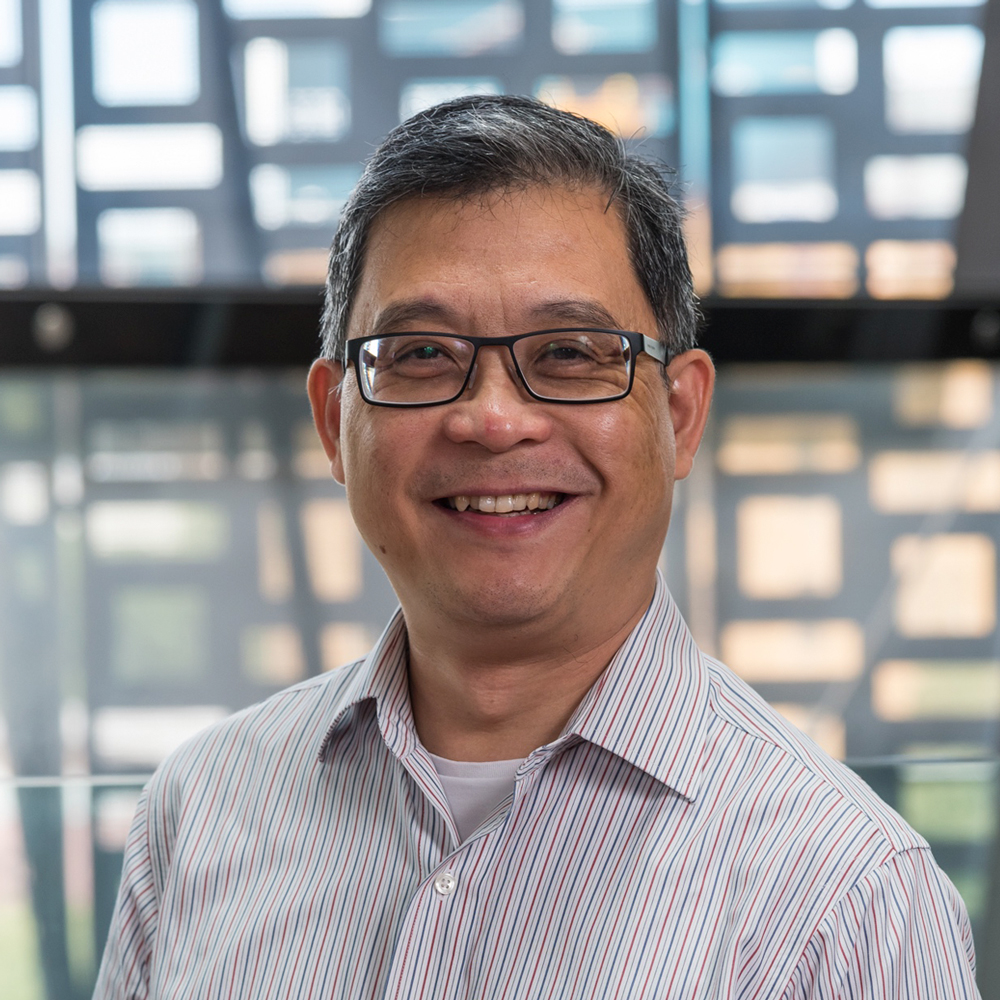}}]{Xu Guandong} is Chair Professor of Artificial Intelligence, Director of University Research Facility of Data Science and Artificial Intelligence, Director of the Centre for Learning, Teaching and Technology at Education University of Hong Kong. His innovative research has earned him many international acclaims. With a penchant for pushing the boundaries of knowledge, he is a trailblazer in applying disruptive technologies to drive research excellence and practise innovation across multi-disciplinary realms. His work encompasses a wide spectrum of fields, including recommender systems, information retrieval, data-driven knowledge discovery, and social computing, and continues to garner increasing citations from the academic community. He was listed in the top 2\% in the Stanford list of the world’s most-cited scientists for consecutive years. Beyond his research, he also serves as the founding Editor-in-Chief of the Human-centric Intelligent Systems Journal (Springer) and the Assistant Editor-in-Chief of the World Wide Web Journal (Springer). His visionary role as the founding Steering Committee Chair of the International Conference of Behavioural and Social Computing further underscores his dedication to advancing these dynamic fields and nurturing the next generation of scholars.
\end{IEEEbiography}

\begin{IEEEbiography}[{\includegraphics[width=1in,height=1.25in,clip,keepaspectratio]{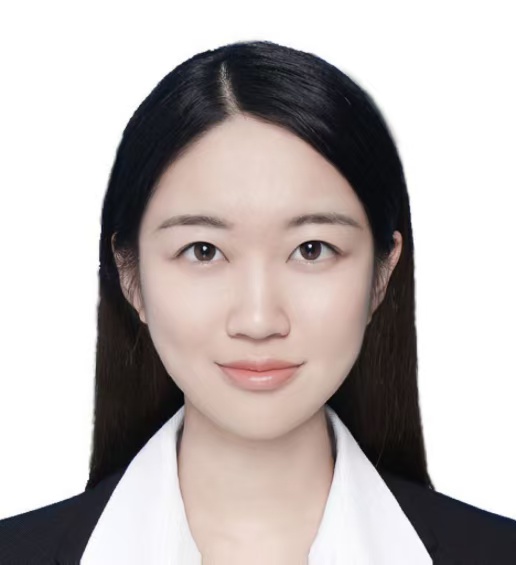}}]{Wanyu Wang}
Wanyu Wang is currently a Ph.D. student in the Department of Information Systems at City University of Hong Kong. Her research interests include recommender systems and large language models. She has published over 20 papers in top-tier conferences such as KDD, WWW, SIGIR, AAAI, CIKM, RecSys, etc.
\end{IEEEbiography}

\begin{IEEEbiography} [{\includegraphics[width=1in,height=1.25in,clip,keepaspectratio]{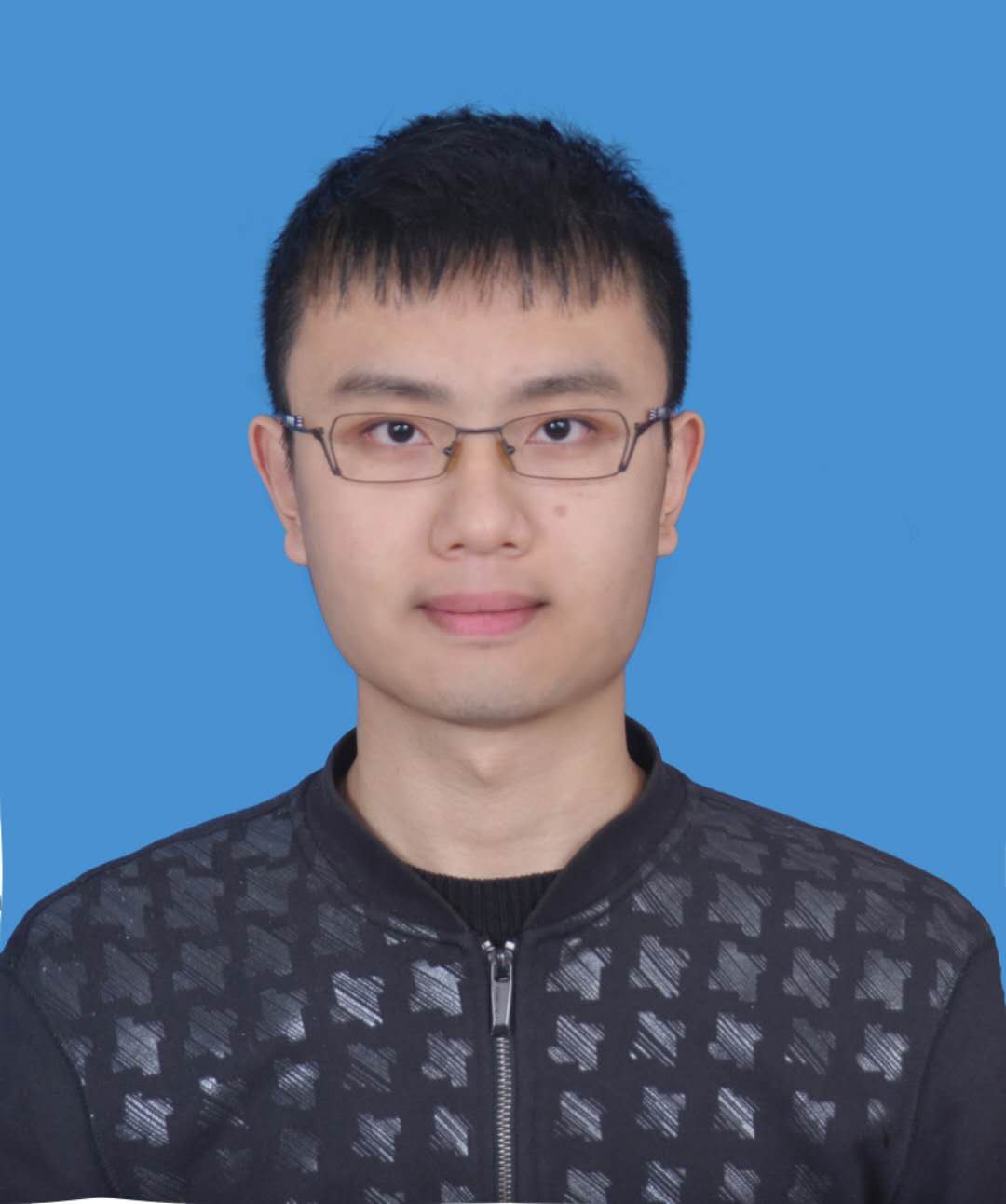}}]{Kaixiang Yang} received the B.S. degree and M.S. degree from the University of Electronic Science and Technology of China and Harbin Institute of Technology, China, in 2012 and 2015, respectively, and the Ph.D. degree from the School of Computer Science and Engineering, South China University of Technology, China, in 2020.

He has been a Research Engineer with the 7th Research Institute, China Electronics Technology Group Corporation, Guangzhou, China, from 2015 to 2017, and has been a postdoctoral researcher with Zhejiang University from 2021 to 2023. He is currently an associate professor with the School of Computer Science and Engineering, South China University of Technology, Guangzhou. His research interests include pattern recognition, machine learning, and industrial data intelligence.
\end{IEEEbiography}

\end{document}